\documentclass[preprint,aps,superscriptaddress,nofootinbib]{revtex4}

\usepackage{graphicx}% Include figure files
\usepackage{dcolumn}% Align table columns on decimal point
\usepackage{bm}% bold math

\makeatletter
\def\simleq{\mathrel{\mathpalette\gl@align<}}
\def\simgeq{\mathrel{\mathpalette\gl@align>}}
\def\gl@align#1#2{\lower.6ex\vbox{\baselineskip\z@skip\lineskip\z@
	\ialign{$\m@th#1\hfill##\hfil$\crcr#2\crcr\sim\crcr}}}
\makeatother
\newcommand{\bec}[1]{\mbox{\boldmath $#1$}}
\newcommand{\fslash}[1]{\ooalign{\hfil/\hfil\crcr$#1$}}
\newcommand{\bra}{\langle}
\newcommand{\ket}{\rangle}
\newcommand{\braket}[1]{\bra #1 \ket}
\newcommand{\qq}{\braket{\bar{q}q}}
\newcommand{\uu}{\braket{\bar{u}u}}
\newcommand{\sbs}{\braket{\bar{s}s}}
\newcommand{\GG}{\braket{\frac{\alpha_s}{\pi} G_{\mu\nu}G^{\mu\nu}}}
\newcommand{\pdiff}{\partial}
\newcommand{\fsl}[1]{\fslash{#1}}

\newcommand{\qGq}{\braket{\bar{q}g_s \sigma\cdot G q}}
\newcommand{\uGu}{\braket{\bar{u}g_s \sigma\cdot G u}}
\newcommand{\sGs}{\braket{\bar{s}g_s \sigma\cdot G s}}
\newcommand{\coeffa}{\frac{\qq}{f_\pi}}

\newcommand{\nn}{\nonumber\\}

  \def\vec#1{\mbox{\boldmath $#1$}}
  \def\gsl#1{\rlap{\slash}#1} 
  \def\Eq.#1{Eq.~(\ref{#1})}
  \def\Disc.{{\rm Im}}
  \def\Cont.{{\rm Re}}
  \def\G{\gamma} 
  \def\VF{\Gamma_+} 
  \def\<>#1{\langle{#1}\rangle} 
  
  \def\d{\bar d} 
  \def\u{\bar u}

  \def\q{\gsl q}

  \def\MB{M} 
  \def\MBS{\MB^2} 
  \def\MBQ{\MB^4} 
  \def\MS{m_\Sigma} 
  \def\ML{m_\Lambda} 
  \def\ES{E_\Sigma} 
  \def\EL{E_\Lambda} 
  \def\dM{\Delta m} 
  \def\m{m_\pi} 
  \def\even{{\rm E}} 
  \def\odd{{\rm O}}
  \def\lL{\lambda_\Lambda}
  \def\lS{\lambda_\Sigma}
  \def\LL{\lL\lS}

\def\sth{\sqrt{s_{th}}}
\def\gs{g_{\Sigma}}
\def\gl{g_{\Lambda\pi}}

\begin{document}

%\preprint{APS/123-QED}

\title{Meson-Baryon Couplings from QCD Sum Rules}% Force line breaks with \\
\thanks{This paper is dedicated to the memory of Dr. Hidenaga Yamagishi.}
		
\author{Takumi Doi}%
\affiliation{%
Department of Physics, Tokyo Institute of Technology, Meguro, Tokyo 152-8551, Japan
}%

\author{Yoshihiko Kondo}%
\affiliation{%
Kokugakuin University, Higashi Shibuya, Tokyo 150-8440, Japan
}%

\author{Makoto Oka}%
\email[email: ]{oka@th.phys.titech.ac.jp}
\affiliation{%
Department of Physics, Tokyo Institute of Technology, Meguro, Tokyo 152-8551, Japan
}%

\date{\today}% It is always \today, today,
             %  but any date may be explicitly specified

\begin{abstract}
Coupling constants of the pseudoscalar mesons to the octet baryons are calculated 
in the QCD sum rule approach.  
Two-point correlation function of the baryons are evaluated in a single meson state 
and the vacuum, which yields the designated coupling. 
The emphasis is on the flavor SU(3) structure of the coupling constants and reliability in 
extracting the coupling constants from the two-point correlation functions.
We first calculate the baryon-diagonal couplings and study the reliability of the sum rule. 
The $F/D$ ratio of the coupling is determined in the SU(3) limit.
We further formulate the baryon-off-diagonal couplings using the projected correlation functions and the vertex functions,
so that the unwanted excited states do not contaminate the sum rule.
As an example, the $\pi\Lambda\Sigma$ coupling constant is calculated and the flavor SU(3) 
breaking effect is studied.
We find that the effect of SU(3) breaking on the $\pi\Lambda\Sigma$ coupling constant is 
small.
\end{abstract} 

\maketitle    

\section{introduction}

Quantum chromodynamics (QCD) has overwhelming evidences as the right theory of strong 
interactions of quarks and gluons.  It successfully describes hard processes of hadrons, such 
as the scaling behaviors of structure functions, and the positron-electron annihilation cross 
sections \cite{PDG-QCD}.  The lattice QCD gives nonperturbative properties of the vacuum, such as color 
confinement, quark condensate and so on, which are consistent with properties of low lying 
hadrons.  Some (but not all) properties of hadrons, i.e., masses, are now available directly 
from lattice QCD \cite{LQCD-conf}.  QCD also suggests fascinating possibility of phase transitions of the 
vacuum at finite temperature and/or baryon density \cite{QCD-finiteT}.  
A dedicated accelerator has just begun to look for evidences of such phase transition \cite{RHIC}.

On the other hand, applications of QCD to rich phenomena of low energy hadrons are 
not in bloom yet.  So far, hadronic interactions have been studied mostly in phenomenological 
approaches.  The most prominent example is the nuclear force.  State-of-art $N-N$ 
potentials, which are valid to relatively high energy ($\sim 1$ GeV) , are available in the 
market \cite{NNpot,Nijmegen}.  
They are based on the meson exchange picture, which has been developed in these 
fifty years since Yukawa proposed the pion exchange interaction \cite{Yukawa}, and phenomenological 
short-range part, which should be attributed to the dynamics of quarks and gluons inside the 
baryon \cite{QCM-ptp}.   Although it is highly desirable to establish a foundation of such potentials from the 
QCD viewpoint, no immediate resolution is expected at this moment.

Recent development of hypernuclear physics has lead us to the level that the interactions of 
hyperons, $\Lambda$, $\Sigma$, $\Xi$, and so on (collectively denoted by $Y$), can be 
fairly well studied from the observed spectra of hypernuclei as well as  the data from hyperon 
productions and reactions \cite{Hyper2000}.  
It has been found that the $\Lambda-N$ interaction is somewhat 
weaker than the $N-N$ interaction, and its spin dependent part has new features \cite{ALS-oka}.  
The $Y-N$ interaction is naturally regarded as a generalized nuclear force by including the third flavor, strangeness \cite{YN-SU3}.  Thus, we consider a larger flavor space, i.e., the flavor SU(3).  In terms of SU(3) , the systems composed of two octet baryons belong to 
$$ 8\times 8 = 1+8_s+27 + 8_a+10 +\overline{10} $$
irreducible representations.  Among them the first (last) three are symmetric (antisymmetric) under the exchange of two baryons, and therefore appear in the channels with antisymmetric (symmetric) spin-orbital combination.  The $N-N$ systems have the hypercharge $Y=B+S=+2$ and thus belong  to either 27 or $\overline{10}$ representation.  In other words, all the other representations are not accessible by $N-N$, but can be reachable only by $Y-N$ and $Y-Y$ interactions.  In this sense, study of the $Y-N$ and $Y-Y$ interactions is important for the complete understanding of the baryonic interactions.

Theoretical approaches to the $Y-N$ interaction are naturally to generalize the picture of the $N-N$ interaction by considering exchanges of the pseudoscalar octet, the vector octet  as well as the singlet pseudoscalar and vector mesons.  Again the short-range parts of the potential are given phenomenologically. Much efforts have been made to analyze experimental data to draw a consistent picture of the hyperon interactions. Yet the the most fundamental quantities, i.e., the meson-baryon coupling constants, which are essential in constructing the meson exchange forces of baryons, have been treated as unknown parameters.  In order to reduce the ambiguity,  the SU(3) relations of the coupling constants are often employed, which leaves a free parameter, i.e., the $F/D$ ratio.  A very popular phenomenological $Y-N$ potential is 
the Nijmegen model \cite{Nijmegen}, which has several different versions. 
There the $F/D$ ratio is treated as a free parameter and varies among the different versions. 
However, the effect of SU(3) breaking may well be sizable considering a wide variety of meson masses within the octet \cite{Lutz}.

Under these circumstances, it is highly desirable to calculate the meson-baryon coupling constants from the fundamental theory.  What is the $F/D$ ratio in the SU(3) limit?
How is the SU(3) symmetry broken at the meson-baryon vertices?
A lattice QCD calculation may be preferable, but at this moment is still preliminary \cite{LQCDpiNN}.
Thus alternative analytic approaches may be useful to analyze the SU(3) symmetry structure of the couplings.

In this paper, we employ QCD sum rule approach to the coupling constants of  the pseudoscalar octet mesons and the octet baryons.  
The QCD sum rule \cite{SVZ,RRY} is generally a relation derived from 
a correlation function in QCD and its analytic property.
The correlation function is calculated by the use of operator product expansion (OPE) in the deeply Euclidean region 
on one hand, and is compared with that calculated for a phenomenological parameterization.
The sum rules relate hadron properties directly to the QCD vacuum condensates as well as the other fundamental constants.  Most applications consider two-point correlation functions of hadronic interpolating operators, and derive relations of masses and other single-particle properties 
of the designated hadron (ex.~\cite{Ioffe}).  It has been applied to the properties of hadrons at finite temperature and density \cite{QCDSR-finiteTrho}, 
as well as the calculation of the scattering lengths of two hadrons \cite{QCDSRNN}.  
The obtained sum rules generally give predictions up to a few tens of per cent ambiguity, 
which may come from ambiguities of the QCD parameters as well as 
higher orders in OPE truncation and contaminations from excited  and continuum states. 

Application of the sum rule to the meson-baryon coupling constants, (mostly  the $\pi N N$ coupling constant), was started by Reinders et al. {\cite{Reinders, RRY2}}, who pointed out that the use of three point functions results in $g_{\pi NN}$ which is inconsistent with the  Goldberger-Treiman (GT) relation.  They then considered two-point correlation function with the pion in the initial state, 
\begin{equation}
\Pi^{\alpha\beta}  (q, p)  =  i\int d^4x\, e^{iq\cdot x} \langle 0| 
{\rm T}\left[ J^{\alpha}_N(x) \bar J^{\beta}_N(0)\right]
|\pi(p)\rangle
\label{eq:corr-pi}
\end{equation}
and showed that the sum rule for the first nonperturbative term in OPE gives the GT relation with $g_A=1$.  
Later, Shiomi and Hatsuda {\cite{SH}} improved the sum rule in the soft-pion limit ($p\to 0$) including higher orders in the OPE.  However, Birse and Krippa {\cite{BK1}}  pointed out that the sum rule at $p\to 0$ simply 
reflects the result of the GT relation and does not constitute an independent sum rule from 
that for the nucleon mass. 
It can be easily shown that the correlation function (\ref{eq:corr-pi}) is related to the correlation function 
without the pion in the initial state using the soft pion reduction formula.  
Therefore in order to obtain an independent sum rule, we need to take into account finite pion momentum $p$. 

As the correlation function (\ref{eq:corr-pi}) has the Dirac indices ($\alpha, \beta$), it has several independent terms, which give independent sum rules.  Kim, Lee and Oka {\cite{KLO}} investigated dependencies 
on the Dirac structure and suggested that the tensor (T) structure, proportional to 
$\gamma_5 \sigma^{\mu\nu} q_{\mu} p_{\nu}$, 
gives the most reliable result.  This conclusion was further refined by 
Kim, Doi, Oka and Lee {\cite{KD1, DK1}}.
The latter  also  checked dependencies on the choice of the interpolating field operator for 
the nucleon. They found that this point is especially important in deriving the $F/D$ ratio of the coupling constants, when the formulation is extended to the SU(3) case.
Recently, Kondo and Morimatsu proposed a novel construction of the sum rule 
using projected two-point correlation functions \cite{KM1,KM2}, and 
showed that the coupling constant can be defined without the ambiguity from the choice 
of the effective interaction Lagrangian.

\medskip
The organization of this article is as follows. 
We first review the formulation of the QCD sum rule for the meson-baryon coupling constant
in section~\ref{sec:formalism},  summarizing the arguments given in Refs.~\cite{DK1,DK2}.
In section~\ref{sec:mbb-coupling},
we formulate the sum rules for the ``baryon-diagonal'' coupling constants, such as
$\pi NN$, $\eta NN$, $\pi \Sigma \Sigma$ and so on,  and  analyze their SU(3) structure. 
In the SU(3) limit, these couplings are related with each other.
More specifically, the ratios of any two couplings are parametrized 
by a single common parameter, the $F/D$ ratio.
As the $F/D$ ratio parametrize two allowed octet combinations in
the coupling, it cannot be determined by the SU(3) symmetry alone, while
requiring a larger symmetry, such as the SU(6) symmetry 
of the nonrelativistic quark model, determines the ratio uniquely.

In the potential model approaches~\cite{Nijmegen}, 
the $F/D$ ratio for each exchanged meson octet is considered as a free parameter,
although it is often fixed to the value taken from the SU(6) symmetry, or the value
determined by the semileptonic decays of hyperons.
The latter provides us with the $F/D$ ratio for the axialvector charges of the baryons,
and thus one has to rely on the GT relation to 
apply it to the $F/D$ ratio of the coupling constants.
Clearly it is desirable to determine the $F/D$ ratio of the meson-baryon coupling constants
directly from QCD only with the constraint of the SU(3) symmetry.
Moreover, a careful study of the applicability of the sum rules in the SU(3) limit will give us 
confidence on the sum rule analysis when the SU(3) symmetry is broken, which
is a subject of section~\ref{sec:mbb'-coupling}.

The projected two-point correlation functions, proposed in Refs.~\cite{KM1,KM2}
can be applied to general meson-baryon couplings including the baryon-off-diagonal cases.  
In section~\ref{sec:mbb'-coupling}, we take the $\pi\Lambda\Sigma$ coupling as an example 
and  explain the projected correlation method and evaluate the sum rule for the 
$\pi\Lambda\Sigma$ coupling constant.  We also study effects of the SU(3) breaking on 
this coupling constant.
 
We give a conclusion in section~\ref{sec:conclusion}.

\section{Sum rules for meson-baryon coupling constants}
\label{sec:formalism}

Reinders, Rubinstein and Yazaki, in their celebrated Physics Reports article{\cite{RRY}}, 
summarized their pioneering works on the QCD sum rules.  They studied how the masses of
various mesons and baryons are determined directly from QCD with the help of dispersion
relation which connects operator product expansion (OPE) of QCD operators in the deep 
Euclidean region, which gives the OPE side of the sum rule, 
and the realistic spectral function at the on-mass-shell region, which gives the phenomenological side.
The OPE side contains information of nonperturbative nature of QCD in terms of matrix 
elements, or condensates, of various local operators, such as $\qq$ and $\GG$.

They also considered how the $\pi NN$ coupling constant is calculated in the QCD
sum rule approach.  They studied two correlation functions for the sum rule, 
the three-point correlation function~\cite{Reinders,RRY2},
\begin{eqnarray}
\Pi (q,q',p) = \int d^4x e^{iq'x}\int d^4y e^{-ipy}
\braket{0| {\rm T}[J_{B_1}(x) J_m (y) \bar{J}_{B_2}(0)] |0},
\label{eq:3point_corr}
\end{eqnarray}
and the two-point correlation function with an external meson 
field~\cite{Reinders},
\begin{eqnarray}
\Pi(q,p) = i\int d^4x\ e^{iq\cdot x} \bra 0|{\rm T}[J_{B_1}(x)\bar{J}_{B_2}(0)] 
|m(p)\ket ,
\label{eq:2point_corr}
\end{eqnarray}
where $p$ is the momentum of the meson $m$.
$J_{B_1}$ and $J_{B_2}$ denote local operators which interpolate 
the QCD vacuum and the baryon states, $B_1$ and $B_2$, respectively.
$J_m$ is a similar operator for the meson $m$.
They are called either interpolating field or ``current''.

In principle, both the above correlation functions can be used to construct a sum rule for
the coupling constant.
However, in practice, the three-point correlation function has disadvantages
that one has to assume the meson pole dominance
and neglect the contribution from higher excited states.
This assumption is not justified because the OPE is valid only
in the deep Euclidean region of the meson momentum ($p^2 \rightarrow -\infty$).
It was demonstrated, indeed, by Maltman~\cite{maltman} that
the sum rule for the $\pi NN$ coupling has large contamination 
from the excited pions, $\pi (1300)$ and $\pi (1800)$. 
On the other hand, for the two-point correlation function,
we may take into account the relevant meson exclusively via the corresponding 
meson matrix elements.
Kim \cite{Kim1} further pointed out that the double dispersion relation used in analytic continuation
of the three point function needs a special care in order to keep the consistency with the soft-pion
limit, and concluded that, in the soft-pion limit, the sum rule from 
the three-point correlation function is reduced to that from the two-point correlation function.
Therefore, we adopt the two-point correlation function in the present analyses.

The phenomenological side of Eq.~(\ref{eq:2point_corr}) contains not only the pole term
which describes the designated meson-baryon coupling, 
\begin{eqnarray}
\Pi(q,p) \sim g_{mB_1 B_2} \cdot
\frac{1}{(q^2 - m_{B_1}^2)}\ \frac{1}{\{ (q-p)^2 - m_{B_2}^2 \}} .
\label{eq:mbbpole}
\end{eqnarray}
but also the terms which represent the transitions from the ground state baryon
to excited baryon resonances as well as
the contribution from the intermediate states due to the meson-baryon scatterings.
We call them ``continuum'' contribution hereafter.

Generally speaking, it is difficult to eliminate these unwanted contributions
and extract only the aimed pole term, Eq.~(\ref{eq:mbbpole}).
However, in the case of the baryon-diagonal couplings 
($B_1 = B_2 \equiv B$, $g_{mB_1 B_2} \equiv g_{mB}$)
it is possible to single out the $g_{mB}$ term in the soft-meson approximation.
Namely, after performing the Taylor expansion with respect to the meson momentum $p$,
the contribution from the ground state has a double pole structure as
\begin{eqnarray}
\Pi(q,p) \sim g_{mB} 
\frac{1}{(q^2 - m_{B}^2)^2}.
\end{eqnarray}
On the other hand, terms from 
the excitations of baryon resonances remain as a single pole structure,
\begin{eqnarray}
\sim g_{mB B^*} \cdot
\frac{1}{(q^2 - m_{B}^2)}\ \frac{1}{(q^2 - m_{B^*}^2)} .
\end{eqnarray}

Similarly, the intermediate meson-baryon scatterings give 
a single pole structure \cite{KM1}.
Therefore, by extracting the double pole term exclusively, 
we can evaluate $g_{mB}$ without contamination.
The concrete procedure will be shown in section~\ref{sec:mbb-coupling}.

The above technique is valid only when the masses of $B_1$ and $B_2$ are equal, or
close enough compared to the baryon excitation energy.
We need further elaboration for the off-diagonal couplings, which 
will be discussed in section~\ref{sec:mbb'-coupling}.

In the two-point correlation function, Eq.~(\ref{eq:2point_corr}), we have suppressed 
the Dirac indices coming from the interpolating field.
The Dirac index structure of $\Pi(q,p) \equiv \Pi^{\alpha\beta}(q,p)$ can be 
written as a sum of four independent terms allowed by the Lorentz covariance, 
\begin{eqnarray}
\Pi (q,p) &=&
  i\gamma _5 \fslash{p}\ \Pi ^{\rm PV}
+ i\gamma _5\ \Pi ^{\rm PS}
+ \gamma _5 \sigma ^{\mu\nu} q_\mu p_\nu\ \Pi ^{\rm T}
+ i\gamma _5 \fslash{q}\ \tilde{\Pi}^{\rm PV}.
\label{eq:4Dirac}
\end{eqnarray}
Each of the four Lorentz-scalar correlation functions, $\Pi ^{\rm PV}$, $\Pi ^{\rm PS}$, 
$\Pi ^{\rm T}$ and $\tilde{\Pi}^{\rm PV}$, can be used to construct 
a sum rule, and should give the same result in principle.
In reality, however, they often give different and sometimes 
contradictory results with each other.

Shiomi and Hatsuda~\cite{SH} calculated $\pi NN$ coupling
in the soft-pion limit ($p_\mu\rightarrow 0$), which leaves only $\Pi ^{\rm PS}$, 
the $i\gamma_5$ structure at the order of ${\cal O}((p)^0)$.
Birse and Krippa pointed out that the sum rule from $\Pi ^{\rm PS}$ is 
reduced to the sum rule for the nucleon mass by a chiral rotation.
Thus they argued that
it is necessary to go beyond the soft-meson limit to construct
an independent sum rule for the coupling.
With this remark, they adopted $\Pi ^{\rm PV}$ in their analysis.

Later, Kim, Lee and Oka~\cite{KLO,Kim2} elaborated the analysis from
the viewpoint of the interaction Lagrangian dependence.
They pointed out that the popular choices of the $mB_1B_2$ effective interaction 
Lagrangian, the pseudoscalar coupling
\begin{eqnarray}
{\cal L}_{\rm int}^{\rm PS} &=& 
g_{mB_1B_2} \bar{\psi}_{B_1} i\gamma_5 \phi_m\psi_{B_2} \ ,
\end{eqnarray}
and the pseudovector coupling
\begin{eqnarray}
{\cal L}_{\rm int}^{\rm PV} &=& -
\frac{g_{m{B_1}{B_2}}}{m_{B_1}+m_{B_2}} 
\bar{\psi}_{{B_1}}\gamma_\mu \gamma_5 (\pdiff^\mu\phi_m)\psi_{B_2} ,
\end{eqnarray}
should give the same result when the baryons are taken on their mass-shell states.
In fact, the sum rule relies on the analytic continuation from far off-mass-shell kinematics, 
which is not guaranteed automatically.  They found that the double pole term in the
two-point correlation function safely satisfies the equivalence condition. 
From this observation, in Ref.~\cite{KLO}, they pointed out that 
$\tilde{\Pi}^{\rm PV}$ sum rule is not appropriate
because this structure appears only when we choose the pseudovector effective
interaction Lagrangian.
Later, Kondo and Morimatsu \cite{KM1} pointed out that defining the coupling constant 
by using the vertex function will avoid the ambiguity, which is consistent with
that calculated from the double pole term.

\section{Diagonal meson-baryon couplings and the SU(3) limit}
\label{sec:mbb-coupling}

\subsection{Construction of the sum rules}
\label{subsec:mbb-construction}

We first consider the diagonal meson-baryon couplings, and construct
the coupling sum rules for general baryon interpolating fields \cite{DK1,DK2}.
We choose the baryons, $N$, $\Sigma$ and  $\Xi$,
and the mesons, $\pi$ and $\eta_8$.
As we are interested in the SU(3) symmetry of the coupling constants, we choose 
the baryon interpolating fields so as to form an SU(3) octet.
\begin{eqnarray}
J_N(x;t) &=& \ \ 2\epsilon _{abc}
[\ (u_a^T(x) C d_b(x))\gamma_5 u_c(x) 
+ t\ (u_a^T(x) C\gamma_5 d_b(x))u_c(x)\ ]\ , \nonumber\\
J_\Xi(x;t) &=& -2\epsilon _{abc}
[\ (s_a^T(x) C u_b(x))\gamma_5 s_c(x) 
+ t\ (s_a^T(x) C\gamma_5 u_b(x))s_c(x)\ ]\ , \nonumber\\
J_\Sigma(x;t) &=& \ \ 2\epsilon _{abc}
[\ (u_a^T(x) C s_b(x))\gamma_5 u_c(x) 
+ t\ (u_a^T(x) C\gamma_5 s_b(x))u_c(x)\ ]\ . 
\label{eq:current}
\end{eqnarray}
Here, $a,b,c$ are color indices, $T$ denotes the transpose 
with respect to the Dirac indices, and $C= i\gamma^2\gamma^0$.
We note that there exist two independent interpolating fields for each octet baryon, 
and thus the general interpolating field is a linear combination of the two designated by 
a real parameter $t$ (We will also use $\theta \equiv \tan^{-1} t$ later).
Note that if we restrict ourselves not to use derivatives in the interpolating fields, these
are the most general forms.
The choice $t=-1$, often called the Ioffe current~\cite{Ioffe}, is regarded as a good
interpolating field operator for the ground state baryon.  We, however,
study the sum rule for general $t$ and will see that
the $t$ dependence is a good check point on the validity of the sum rule.

Our main purpose here is to determine the pertinent Dirac structure
\cite{KD1,DK1}.
We set the following criteria:
(1) The result should be independent of the choice of the baryon interpolating field.
(2) The choice minimizes the single pole terms, which come from excited baryon 
resonances and meson-baryon scatterings.
(3) It should not depend on the coupling scheme of the effective Lagrangian.  
%The last criterion is cleared when 
%we take the double pole term in the sum rule, except for the $\tilde{\Pi}^{\rm PV}$ 
%sum rule, in which no double pole term appears.

In going beyond the soft-meson limit, 
we consider three distinct Dirac structures: 
$\Pi^{\rm PV}$ (PV),
$\Pi^{\rm PS}$ (PS), and 
$\Pi^{\rm T}$ (T).
For the PV and T structures, 
we construct the sum rules to ${\cal O}(p)$.
At this order, the terms proportional to the quark mass $m_q$ should not be included in the OPE,
because $m_q$ is of the order $m_\pi^2$ as is seen from
the Gell-Mann--Oakes--Renner relation,
\begin{eqnarray}
-2m_q\qq = m_\pi^2 f_\pi^2 \ .
\end{eqnarray}
The ${\cal O}(p)$ calculation for the PS structure is not useful,
because the OPE is essentially the same as the ${\cal O}(1)$ terms of the same
structure, which are equivalent to the sum rule for the baryon mass.
Therefore, we construct PS sum rule at the order $p^2=m_\pi^2$~\cite{Kim2}.
Note that the terms linear in quark mass $(m_q)$ should be included in the OPE 
at this chiral order.

\begin{figure}[hbtp]
 \caption{ The OPE diagrams considered in this work.
The solid lines denote quark propagators and the wavy lines denote 
gluon propagators. The blob in each figure denotes the matrix
element, or the wave function of
the external meson specified by the dashed line.
}
\begin{center}
\includegraphics[scale=0.5]{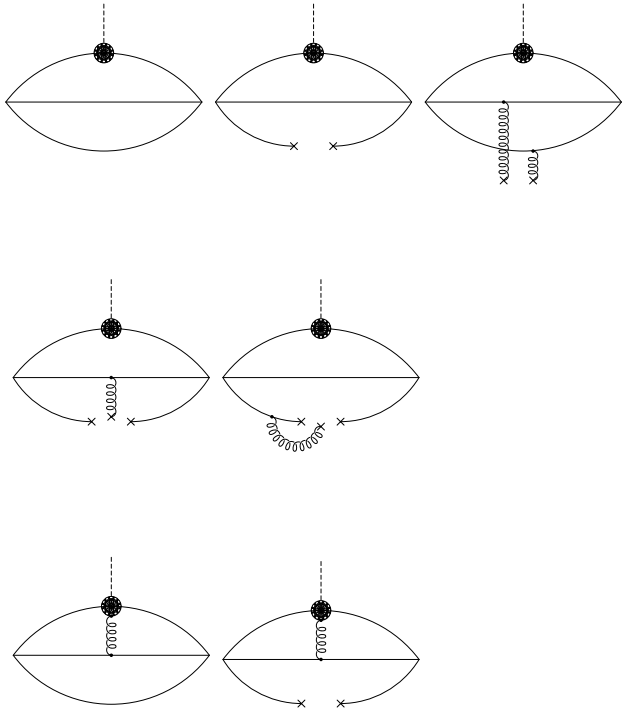}
\end{center}
\label{fig:mb_OPE}
\end{figure}

We calculate the OPE up to dimension 8 for the PS sum rule, and
up to dimension 7 for the PV and T sum rules.
The OPE diagrams which we consider are shown in Fig.\ref{fig:mb_OPE}.
In the OPE calculation, we expand the meson matrix elements
in terms of the momentum of the meson $p$.
Technical details on the OPE calculation are given in Ref.~\cite{DK1}.

%%%%%%%%%  figure %%%%%%%%%%%%%%%
%%%%%%%%%%%%%%%%%%%%%%%%%%%%%%%%

In order to construct the phenomenological side of the sum rule, we adopt 
the effective interaction Lagrangian with the 
PS type coupling. We obtain the correlation function as
\begin{eqnarray}
&\mbox{PV structure}&\ \mbox{at the order}\ {\cal O}(p)
\qquad 
-i\gamma_5\fslash{p} 
\frac{g_{m B}\lambda_B^2(t)m_B}{(q^2-m_B^2)^2}, \\
&\mbox{PS structure}&\ \mbox{at the order}\ {\cal O}(p^2)
\qquad 
i\gamma_5
p^2 \frac{g_{m B}\lambda_B^2(t)}{(q^2-m_B^2)^2}, \\
&\mbox{T structure}&\ \mbox{at the order}\ {\cal O}(p)
\qquad 
\gamma_5\sigma_{\mu\nu}q^\mu p^\nu 
\frac{g_{m B}\lambda_B^2(t)}{(q^2-m_B^2)^2},
\end{eqnarray}
where $\lambda_B(t)$ is the coupling strength of the single baryon state to
the corresponding baryon interpolating field operator $J_B(t)$ defined by
\begin{eqnarray}
\braket{0|J_B(0; t)|B(p)} &=& \lambda_B(t) u_B(p) ,\nonumber
\end{eqnarray}
and we choose the phase so that $\lambda_B(t)$ is real.
The continuum contributions are treated approximately by assuming the QCD duality
that they match the OPE at above an effective threshold, $\sqrt{s_{th}}$,
which is regarded as a parameter and is determined phenomenologically.

The sum rule is obtained by matching the OPE side with the phenomenological side 
and by taking the Borel transformation in order to improve the predictability.
The obtained sum rules are of the form,
\begin{eqnarray}
g_{mB}\lambda_B^2(t)\left[\ 1+ A_{mB}(t) M^2 \ \right] 
&=& 
%e^{m_B^2/M^2}\cdot
%F^{\mbox{\tiny OPE}}_{\mbox{\tiny\it $m$B}}(M^2;t) 
%\quad \equiv \quad
f^{\mbox{\tiny OPE}}_{\mbox{\tiny\it $m$B}}(M^2;t) 
\label{eq:qsr-formula}
\end{eqnarray}
where  $M$ is the Borel mass and $A_{mB}$ denotes 
the sum of the single pole terms.
Thus, the contamination from the single pole terms can be
eliminated by fitting $f^{\mbox{\tiny OPE}}_{\mbox{\tiny\it $m$B}}$ by a
linear function of $M^2$.
Explicit expressions of  
%$F^{\mbox{\tiny OPE}}_{\mbox{\tiny\it $m$B}} (M^2;t)$ 
$f^{\mbox{\tiny OPE}}_{\mbox{\tiny\it $m$B}} (M^2;t)$ 
are given in Ref.~\cite{DK1}.

\subsection{Pertinent Dirac structure for meson-baryon coupling sum rules}
\label{subsec:Dirac}

We here determine the most pertinent Dirac structure for the meson-baryon coupling sum rule
out of the PV, PS and T structures given above.
First of all, we focus on the sensitivity to the continuum threshold, $\sqrt{s_{th}}$.
We often choose $\sqrt{s_{th}}$ to be the position of the first excited state, 
but examine whether the result is not too sensitive to the choice.
In the present case, we first set the threshold 
$s_{th}=2.07  {\rm GeV}^2$ ($\sqrt{s_{th}} = 1.44 {\rm GeV}$),
that corresponds to the Roper resonance, and $s_{th}$ is changed to
$s_{th}=2.57  {\rm GeV}^2$ ($\sqrt{s_{th}} = 1.60 {\rm GeV}$)
to see how the sum rule is sensitive to the threshold.
We find that the result from the PV sum rule changes by about 15\%
at the Borel mass $M^2 = 1 {\rm GeV}^2$.
We also find that the PV sum rule has a large uncertainty in the linear fitting of 
Eq.~(\ref{eq:qsr-formula}).
On the other hand, the PS and T structures are insensitive to $s_{th}$. At $M^2 = 1 {\rm GeV}^2$, 
the difference is only $2-3 \%$ level, and the uncertainty in the linear fitting is also small. 
Therefore, we conclude that the sum rule from PV structure is not reliable.
Kim et al., in fact, found that this sensitivity of the PV sum rule may be understood by 
coherent addition of the positive and negative parity excited states {\cite{KLO,Kim2}}.

As the next step, we compare the PS and T sum rules, focusing on the dependence of the OPE 
on the baryon interpolating field (i.e. the dependence on $t$).
Ideally, the physical parameter $g_{mB}$ should be independent of $t$, 
but in practice, the truncation of OPE brings some spurious $t$-dependence.
Therefore, the above constraint tells us how the truncated 
OPE calculation is reliable.
In Eq.~(\ref{eq:qsr-formula}), as the right hand side, $f^{\mbox{\tiny OPE}}_{\mbox{\tiny\it $m$B}}$, is 
a quadratic function of $t$, the obtained  $g_{mB}\lambda_B^2(t)$
is also quadratic.
In the SU(3) limit, the strength $\lambda_B$ should be common
to all the octet baryons, and thus a ``good'' sum rule must give 
a common shape (in $t$) for the obtained $g_{mB}\lambda_B^2(t)$.

\begin{table}
\caption{\label{tab:table1}
QCD parameters in the SU(3) limit.}
\begin{ruledtabular}
\begin{tabular}{cccc}
$\qq$  &  $m_0^2 \equiv \qGq / \qq$ 
& $\delta^2$ & $\GG$\\
\hline
$ -(0.23~ {\rm GeV})^3$ & $0.8~{\rm GeV}^2$ & $0.2~{\rm GeV}^2$ & $(0.33~{\rm GeV})^4$  \\
\end{tabular}
\end{ruledtabular}
\end{table}

%%%%%%%%%  figure %%%%%%%%%%%%%%%
\begin{figure}[hbtp]
%\begin{center}
 \caption{
  $g_{mB}\lambda_B^2(t)$
  from the T sum rule is plotted as a function of $t$, for the 
  $\pi NN$, $\eta NN$, $\pi\Xi\Xi$, $\eta\Xi\Xi$, 
  $\pi\Sigma\Sigma$, and $\eta\Sigma\Sigma$ couplings.
  We choose the Borel window as $0.65 \leq M^2 \leq 1.24\ {\rm GeV^2}\ $,
  and the continuum threshold as $s_{th} = 2.07\ {\rm GeV^2}$.
}
\begin{center}
\includegraphics*[width=10cm]{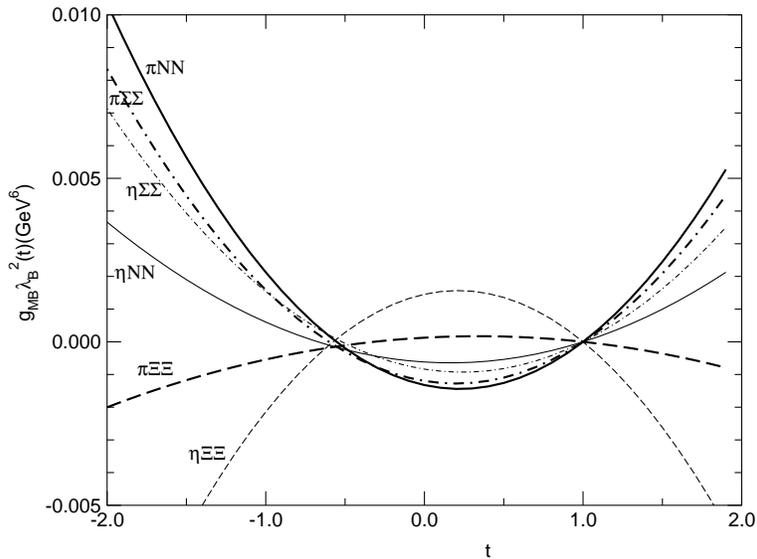}
\end{center}
\label{fig:t_dep-T}
\end{figure}
%%%%%%%%%%%%%%%%%%%%%%%%%%%%%%%%

%%%%%%%%%  figure %%%%%%%%%%%%%%%
\begin{figure}[hbtp]
% \begin{center}
 \caption{
  $g_{mB}\lambda_B^2(t)$
  from the PS sum rule
  is plotted as a function of $t$.
}
\begin{center}
\includegraphics*[width=10cm]{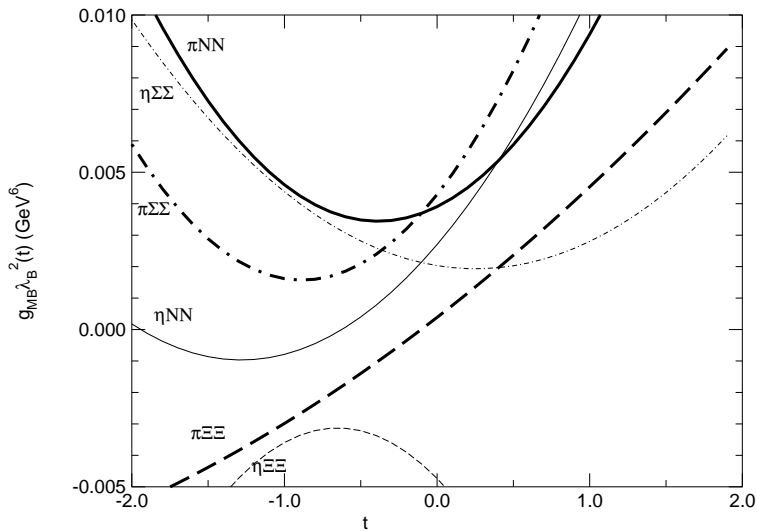}
\end{center}
\label{fig:t_dep-PS}
\end{figure}
%%%%%%%%%%%%%%%%%%%%%%%%%%%%%%%%

Fig.~\ref{fig:t_dep-T} shows $g_{mB}\lambda_B^2(t) $
as a function of $t$ for the T sum rules, while the same for the PS sum rules is
given in Fig.~\ref{fig:t_dep-PS}.
In drawing the curves a standard set of the QCD parameters
and pion matrix elements are employed, 
the values of which are given in Table~\ref{tab:table1}.
Among them, $\delta^2$ is defined in Eq.~(\ref{eq:delta}), in Appendix A,
where the parametrizations of the pion matrix elements are explained.

It is interesting to see that, for the T sum rule, 
(1) all the curves have zeros at $t=1$ and also near $t\sim -0.6$, and 
(2) each curve has an extremum at around $t\sim 0.2$.
One observes that the curves are well proportional to each other, indicating they 
are given by a single (quadratic) function of $t$, i.e., $\lambda_B^2(t)$.
In contrast, the curves for the PS sum rule are random and do not show a single
behavior.
Therefore, we conclude that the T sum rule is more appropriate than the PS sum rule.
This statement is further supported by the fact that
the $t$-dependence of $\lambda_B^2(t)$ from baryon mass sum rules
is consistent with 
that extracted from the T sum rules~\cite{DK1}.

Thus we have seen that the T sum rule passes the first two criteria given 
in sect.~\ref{subsec:mbb-construction}.  In fact, the third criterion is also satisfied by the
T sum rule as it does not depend on the forms of the coupling in the effective Lagrangian \cite{KLO}.
We conclude that the tensor (T) is the most pertinent  Dirac structure.

\subsection{The $F/D$ ratio of the octet meson-baryon couplings%
\protect\footnote{The results presented in this section are modified from the previous calculations by correcting the sign of the $\delta^2$ parameter.  See Appendix A for details.}}
\label{subsec:F/D}

We are now ready to construct sum rules 
from the $\gamma_5 \sigma_{\mu\nu}q^\mu p^\nu$ (T) structure 
in the SU(3) limit and to determine the $F/D$ ratio.
In particular, we investigate the $t$-dependence of the ratio using
the general interpolating fields for the baryons.
We consider the octet baryons $B^a$, and the octet mesons $m^a$,
where $a = 1\ldots 8$ denotes the flavor index for the octet representation.
We note that in the SU(3) limit there is no mixing of the singlet $\eta_1 (=m^0)$ and 
the octet $\eta_8 (=m^8)$.
The $m^{c}B^{a}B^{b}$ coupling has two independent terms, 
the antisymmetric (F) and symmetric (D) couplings
\begin{eqnarray}
{\cal L}_{F} &=& -F \, if_{abc} \,\bar\psi_{B}^{a} i \gamma^5\psi_{B}^{b}
\phi_m^{c}\nonumber\\
{\cal L}_{D} &=& D \, d_{abc} \,\bar\psi_{B}^{a} i \gamma^5\psi_{B}^{b}
\phi_m^{c}
\end{eqnarray}
where $f$ and $d$ are the group algebraic constants of SU(3).
Then all the couplings are given by these two parameters, or equivalently
in terms of
\begin{equation}
g_{\pi N}= (F+D) \qquad {\rm  and}\qquad  \alpha={F\over F+D}\ .
\end{equation}
The explicit forms of the individual couplings are given by \cite{deSwart} 
\begin{eqnarray}
g_{\eta N} &=& \frac{1}{\sqrt{3}}(4\alpha -1) g_{\pi N}\;; \quad
g_{\pi\Xi} = (2\alpha -1 ) g_{\pi N}\nonumber \ ,\\
g_{\eta\Xi} &=& -\frac{1}{\sqrt{3}}(1+2\alpha ) g_{\pi N}\;;  \quad
g_{\pi\Sigma}= 2\alpha\ g_{\pi N} \nonumber\ ,\\
g_{\eta\Sigma} &=& g_{\pi\Lambda\Sigma} = {2\over \sqrt{3}} (1-\alpha ) g_{\pi N}
\label{eq:su3-deSwart}
\end{eqnarray}

We note that our sum rules satisfy these relations at the level of the OPE expressions.
This is a consequence of using the
baryon interpolating fields constructed according to the SU(3) symmetry.
Hence, it provides the consistency of the sum rules with the SU(3)
relations for the couplings.
We determine $g_{mB}$ for $m = \pi$ or $\eta$
and $B = N, \Xi$ or $\Sigma$ by linear fitting of the right hand side
of Eq.~(\ref{eq:qsr-formula}).
Taking the ratio of any two different coupling constants, we can 
convert it into the $F/D$ ratio
according to Eq.~(\ref{eq:su3-deSwart}).

%%%%%%%%%   figure   %%%%%%%%%%%%%%%%%%%%%
\begin{figure}[btp]
%\begin{center}
 \caption{The $F/D$ ratio from the T sum rules
   is plotted as a function of $\cos\theta$,
   where $\theta$ is defined as $\tan\theta = t$. 
   (See the text.)
   Corresponding $t$ is also shown at the top of the figure.
   Circles are obtained with 
   $0.65 \leq M^2 \leq 1.24\ {\rm GeV^2}$ and the continuum
   threshold $s_{th} = 2.07\ {\rm GeV^2}$, triangles ;
   $0.65 \leq M^2 \leq 1.24\ {\rm GeV^2}$, $s_{th} = 2.57\ {\rm GeV^2}$,
   squares ;
   $0.90 \leq M^2 \leq 1.50\ {\rm GeV^2}$, $s_{th} = 2.07\ {\rm GeV^2}$.
   In the realistic region 
   $-0.75 \protect\simleq \cos\theta \protect\simleq 0.61$,
   the $F/D$ ratio is insensitive to $t$.
}
\begin{center}
\includegraphics*{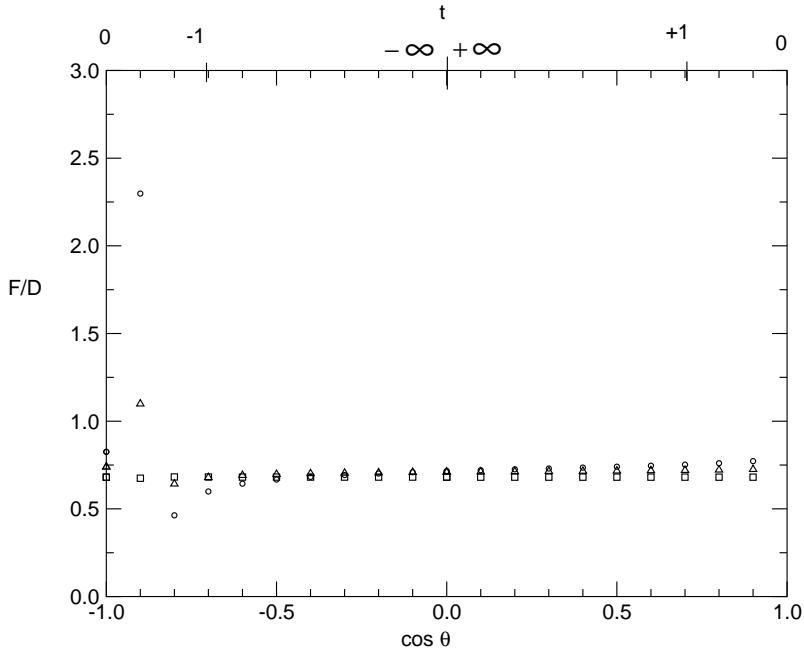}
\end{center}
\label{fig:F/D-T}
\end{figure}
%%%%%%%%%%%%%%%%%%%%%%%%%%

In Fig.~\ref{fig:F/D-T}, the $F/D$ ratio is plotted as a function of 
$\cos\theta$. Here, to investigate the whole range of 
$ -\infty \leq t \leq +\infty $,
we introduce a new parameter $\theta$ defined as $\tan\theta = t$. 
Thus, the range $0 \leq t \leq +\infty $ 
corresponds to $ 0 \leq \theta \leq \pi / 2$ while
the range $ -\infty \leq t \leq 0 $
spans $ \pi / 2 \leq \theta \leq \pi $.
In Fig.~\ref{fig:F/D-T}, circles are obtained from the linear fitting within 
the Borel window $ 0.65 \leq M^2 \leq 1.24\ {\rm GeV^2}$
with 
the continuum threshold $s_{th} = 2.07\ {\rm GeV^2}$.
To see the sensitivity to these parameters, we also
calculate the ratio using
(1) $ 0.65 \leq M^2 \leq 1.24\ {\rm GeV^2}$, $s_{th} = 2.57\ {\rm GeV^2}$ 
(triangles), and 
(2) $ 0.90 \leq M^2 \leq 1.50\ {\rm GeV^2}$, $s_{th} = 2.07\ {\rm GeV^2}$
(squares).

We see that the $F/D$ ratio is insensitive to the continuum threshold, as discussed before,
and is also insensitive to the choice of the Borel window. 
The curve is flat with respect to $t$\  except around the region $t\sim -0.6$ ($\cos\theta \sim -0.9$).
The behavior around $ t\sim -0.6$ can be understood if we remind the $t$ dependence of
the coupling strengths of the baryon states to the interpolating fields, given in
Fig.~\ref{fig:t_dep-T}.  
There one sees that all the curves have a zero around $t\sim -0.6$, which indicates 
poor convergence of the OPE for this $t$.
Thus, we conclude that the $F/D$ ratio
should be obtained from the data outside $t\sim -0.6$.

Here we moderately take the appropriate region as 
(1) $-0.75 \simleq \cos\theta$ ($ t \simleq -0.9$) \
and (2) $\cos\theta \simleq 0.61$ ($1.3 \simleq t$).
The former constraint gives us the minimum value of
$F/D \sim 0.55$, and the latter constraint gives us
the maximum value of $F/D \sim 0.75$.
Thus we obtain the final result, $F/D = 0.65 \pm 0.10$.
We again stress that this is the determination of the $F/D$ ratio directly 
from (the SU(3) symmetric limit of) QCD without any assumption.
Our result is very close to the value predicted
in the SU(6) quark model, $F/D = {2\over 3}$.  
This is surprising if we realize that 
the SU(6) symmetry, based on the nonrelativistic wave functions of 
three quarks in the baryons, looks fairly remote from QCD.
It would be interesting to study whether this agreement has some
deep reason or is just accidental.
It is worth pointing out that our result is consistent with the $F/D$ ratio 
for the axialvector currents, $F/D \sim 0.57$,
which is extracted from the semi-leptonic weak decays of 
the hyperons \cite{ratcliffe}.  It is reasonable because two $F/D$ ratios 
are to be related by the GT relation.

%%%%%%%%%%  figure %%%%%%%%%%%
\begin{figure}[hbtp]
%\begin{center}
 \caption{The $F/D$ ratio from the PS structure 
   is plotted as a function of $\cos\theta$. See the caption of 
   Fig.~\ref{fig:F/D-T} for the explanation of each symbol. 
   The $F/D$ ratio is sensitive to $t$ and  no reliable prediction is 
   obtained.
}
\begin{center}
\includegraphics*{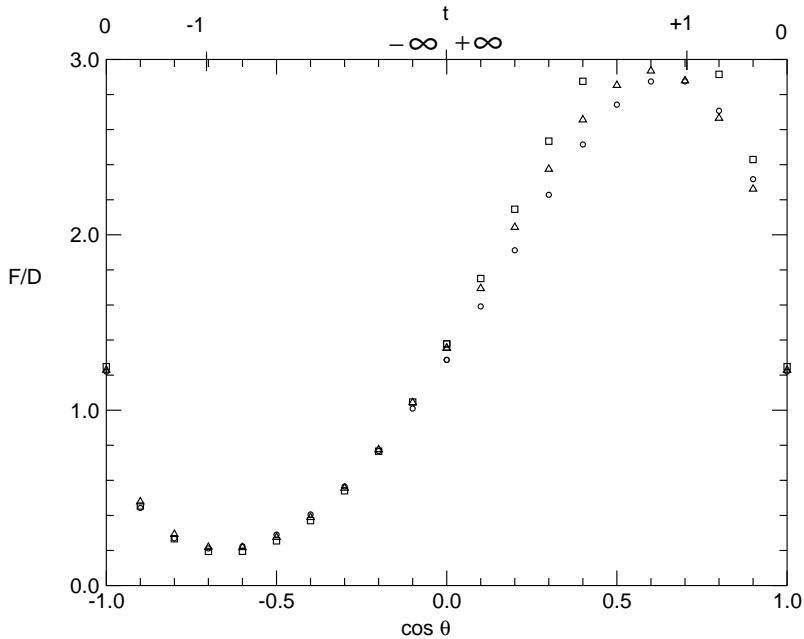}
\end{center}
\label{fig:F/D-PS}
\end{figure}
%%%%%%%%%%%%%%%%%%%%%%%%%%%%%%

Before closing this section, 
let us show the result from the PS sum rule.
In this case too, we can classify the OPE according to 
Eq.~(\ref{eq:su3-deSwart}) and identify the terms responsible
for the $F/D$ ratio.
By taking similar steps as the T sum rules,
we determine the $F/D$ ratio.
Fig.~\ref{fig:F/D-PS} shows the $F/D$ ratio as a function of
$\cos\theta$. Compared with Fig.~\ref{fig:F/D-T},
one sees that 
the $F/D$ ratio is highly sensitive to $t$ and the result is not reliable.
Therefore, we again note that one has to carefully choose
the Dirac structure in the sum rule.

The behavior of the $F/D$ ratio suggests that the PS sum rule is contaminated by 
continuum states, because the peak and the bottom of the curve in Fig.~\ref{fig:F/D-PS}
coincide with those expected from the $F/D$ ratio of the interpolating field operators.
In fact, it is shown \cite{JHO} that $t=1$ ($-1$) corresponds to the pure $F$ ($D$) coupling.
If the sum rule is dominated by the ground state baryon, then it should give a constant $F/D$
ratio, regardless the interpolating field.  The T sum rule is such a case, while the PS sum rule
seems to have significant continuum, which obeys the $F/D$ ratio of the interpolating field
operator. 

\section{Off-Diagonal meson-baryon couplings and the SU(3) breaking effect}
\label{sec:mbb'-coupling}

\subsection{Projected correlation function}
\label{sec:projected}
The double pole structure of the correlation function is found to be crucial in extracting
the coupling constant in the previous section.
It requires a further analysis to generalize the sum rule to off-diagonal meson-baryon couplings, 
where the initial and final baryons have different masses.
In this case, the ground state contribution to the correlation function 
is not given by a double pole, but has single pole structure.
Therefore it is not possible to eliminate contamination from excited states
simply by taking the double pole part of the correlation function, as was done 
in the diagonal couplings.

Recently, Kondo and Morimatsu \cite{KM1} proposed a novel method of extracting
the coupling constant from the two-point correlation function. 
Although they applied their method to the $\pi NN$ coupling constant
in their original work, 
the method is general and is in fact applicable to the off-diagonal case.
Therefore we follow their prescription here, choosing the $\pi \Lambda\Sigma$
coupling as a concrete example to demonstrate the method.

We consider the vacuum-to-pion matrix element of the correlation function 
of the interpolating fields of $\Lambda$ and $\Sigma$:
\begin{eqnarray}\label{Pi}
\Pi(q,p)= i\int d^4xe^{iqx}
\langle 0|{\rm T}[J_{\Sigma}(x)\bar J_\Lambda(0)]|\pi(p)\rangle \, .
\end{eqnarray}
We further define the vertex function by
\begin{eqnarray}\label{Gamma}
\Gamma (q,p) = (\fsl{q} - \MS) \Pi (q,p) (\fsl{q}'- \ML)
\end{eqnarray}
with $q'\equiv q-p$.
Then the $\pi\Lambda\Sigma$ coupling constant, $g\equiv g_{\pi\Lambda\Sigma}$, 
can be defined by the vertex function projected on to the  on-mass-shell baryon states,
\begin{eqnarray}\label{gpiLS}
\u_\Sigma(\vec q) \Gamma (q,p) u_\Lambda(\vec q')|_{q^2=\MS^2,(q')^2=\ML^2}
= i\LL \, g \, \u_\Sigma(\vec q)\G_5u_\Lambda(\vec q'),
\end{eqnarray}
where $u(\vec q)$ denotes the Dirac plane wave spinor (where we suppress the helicity indices 
for simplicity).

Advantage of defining the coupling constant in terms of
the vertex function, $\Gamma$, is that it has no ambiguity originated from the
choice of effective Lagrangian for the coupling.  
Indeed, Eqs. (\ref{Gamma}) and (\ref{gpiLS}) allow us to extract the residue of
the pole term on which both the baryons are on the mass shell.  Such definition is
known to be unique regardless of the form of the effective coupling, such as
the pseudoscalar coupling or pseudovector coupling.
On the other hand, in employing this new approach, we need to compute all possible 
terms with various Dirac structures of the correlation function, as the vertex function
$\Gamma$ is a linear combination of all the terms.

It should be also noted that the kinematical point of the definition Eq.~(\ref{gpiLS}) is an
unphysical point and that the sum rule does not give the coupling constant of that 
point directly. 
However, we will see later that the difference between the coupling constant defined by
Eq.~(\ref{gpiLS}) and the one calculated in the sum rules is small, and therefore can be neglected.

In practical application of the sum rule to the coupling constant, we need a further
elaboration so that the contamination from other poles should be small.
In order to eliminate unwanted contributions from the negative energy solutions, 
Kondo and Morimatsu \cite{KM2} further proposed to use the projected correlation function and 
vertex function defined by
\begin{eqnarray}\label{Pi+}
&&\Pi_+(q,p)=\u_\Sigma(\vec q)\G_0\Pi(q,p)\G_0u_\Lambda(\vec q'),\cr
&&\VF(q,p)=\u_\Sigma(\vec q)(\q-\MS)\Pi(p,q)(\q'-\ML)u_\Lambda(\vec q').
\end{eqnarray}
They satisfy the relation
\begin{eqnarray}\label{Vertex}
\VF(q,p)=(q_0-\ES)(q_0-\EL-\omega_p)\Pi_+(q,p),
\end{eqnarray}
where $\ES=\sqrt{\MS^2+\vec q^2}$, $\EL=\sqrt{\ML^2+\vec q^{'2}}$ and 
$\omega_p=\sqrt{\m^2+\vec p^2}$.
It should be noted that $\Pi_+(q,p)$ has poles at $q_0=\ES$ and $q_0=\EL+\omega_p$ 
but not at $q_0=-\ES$ and $q_0=-(\EL+\omega_p)$.

Regarding $\Pi_+$ and $\Gamma_+$ as functions of the center-of-mass energy, or $q_0$ 
in the reference frame $\vec q=0$, 
the absorptive part of the projected correlation function, $\Disc.\Pi_+$, can be written as
\begin{eqnarray}\label{DiscPi}
\Disc.\Pi_+(q,p)
=&&\pi\delta(q_0-\MS){\Cont.\VF(q,p)\over \EL+\omega_p-\MS}-\pi\delta(q_0-\EL-\omega_p)
{\Cont.\VF(q,p)\over \EL+\omega_p-\MS}
\cr && +\Cont.{1\over (q_0-\MS)(q_0-\EL-\omega_p)}\Disc.\VF(q,p).
\end{eqnarray}
Here we adopt the following notation for the dispersive (continuous) part and 
the absorptive (discontinuous) part, respectively:
\begin{eqnarray}
&&\Cont.F(q)\equiv\lim_{\eta\to 0}{1\over2}\left[F(q)|_{q^0=q^0+i\eta}+F(q)|_{q^0=q^0-i\eta}\right],\cr
&&\Disc.F(q)\equiv\lim_{\eta\to 0}{1\over2i}\left[F(q)|_{q^0=q^0+i\eta}-F(q)|_{q^0=q^0-i\eta}\right].
\end{eqnarray}
In \Eq.{DiscPi}, the first and second terms are the pole terms, which are proportional to the
coupling constant, and the third term is classified as the continuum contribution.
It is noted that all the terms on the right-hand side of \Eq.{DiscPi} are well defined 
in the dispersion integral even in the chiral limit and the flavor SU(3) limit, which 
is another reason to employ the projected correlation function \cite{KM2}.

The dispersion relation for the projected correlation function in the variable $q_0$ is given by
\begin{eqnarray}\label{Dispersion}
\Pi_+(q_0)&=&-{1\over\pi}\int dq_0'
{\Disc.\Pi_+(q_0')\over q_0-q_0'+i\eta}.
\end{eqnarray}
By splitting the projected correlation function into the even and odd parts by 
\begin{eqnarray}
\Pi_{+}^{\even}&=&{1\over2}[\Pi_+(q_0)+\Pi_+(-q_0)] ,\nonumber\\
\Pi_{+}^{\odd}&=&{1\over2q_0}[\Pi_+(q_0)-\Pi_+(-q_0)],
\label{eq:Pievenodd}
\end{eqnarray}
\Eq.{Dispersion} is given by 
\begin{eqnarray}\label{Dispersion2}
\Pi_{+}^{\even}(q_0^2)&=&-{1\over\pi}\int dq_0'
{q_0'\over q_0^2-q_0'^2}\Disc.\Pi_+(q_0'),
\cr
\Pi_{+}^{\odd}(q_0^2)&=&-{1\over\pi}\int dq_0'
{1\over q_0^2-q_0'^2}\Disc.\Pi_+(q_0').
\end{eqnarray}
Applying the Borel transformation, $\hat{\rm L}_M$ defined in the Appendix B, with respect to $q_0^2$, 
we obtain
\begin{eqnarray}\label{BSR}
\hat{\rm L}_M[\Pi_{+}^{\even}]
&=&{1\over\pi}\int dq_0'{q_0'\over\MBS}\exp\left(-{{q_0'}^2\over\MBS}\right)
\Disc.\Pi_+(q_0'),
\cr
\hat{\rm L}_M[\Pi_{+}^{\odd}]
&=&{1\over\pi}\int dq_0'{1\over\MBS}\exp\left(-{{q_0'}^2\over\MBS}\right)
\Disc.\Pi_+(q_0').
\end{eqnarray}
Evaluating the left hand side by the OPE, we obtain the Borel sum rules.

The right-hand side of \Eq.{BSR} is expressed in terms of 
the observed quantities.
We parameterize the absorptive part of the projected correlation function 
for the $\pi\Lambda\Sigma$ vertex:
\begin{eqnarray}\label{Ansatz}
{\rm Im}\Pi_+(q,p)&=& - \u_{\Sigma}(\vec q)i\G_5 u_\Lambda(\vec q')\pi\LL g(q_0,\vec p^2)
\left[{\delta(q_0-\MS)\over q_0-E_\Lambda-\omega_p}
+{\delta(q_0-E_\Lambda-\omega_p)\over q_0-\MS}\right]
\cr&&+\left[\theta(q_0-\sth)+\theta(-q_0-\sth)\right]\Disc.\Pi_+^{\rm OPE}(q,p),
\end{eqnarray}
where $g(q_0,\vec p^2)$ is defined by
\begin{eqnarray}
\Cont.\VF(q,p)= \LL \u_{\Sigma}(\vec q)i\G_5 u_\Lambda(\vec q') g(q_0,\vec p^2).
\end{eqnarray}
In \Eq.{Ansatz} $\sth$ is the effective continuum threshold of the $\pi\Lambda$ or 
$\pi\bar\Lambda$ channel. 
We assume that the asymmetry in the continuum contribution
for the positive and negative energy regions is negligible.

\subsection{$\pi\Lambda\Sigma$ coupling constant}

We evaluate the $\pi\Lambda\Sigma$ coupling constant using the projected sum rule
presented in sect.~\ref{sec:projected}.  
Here we consider only the $t=-1$ (Ioffe) interpolating field for
simplicity.

Substituting \Eq.{Ansatz} into the right-hand side of \Eq.{BSR},
and taking the  limit $\vec p^2\to 0$, we obtain
\begin{eqnarray}\label{dI}
\hat{\rm L}_M(\Pi_{+}^{\even})
&=&\u(\vec q)i\G_5u(\vec q')
{\lL\lS\over\MBS\dM}
\left[\MS\gs \exp\left(-{\MS^2\over\MBS}\right) \right. \nonumber \\
&& \left. \quad -(\ML+\m)\gl \exp\left(-{(\ML+\m)^2\over\MBS}\right)\right]
+({\rm Cont.}),
\end{eqnarray}
for the even part and
\begin{eqnarray}\label{dIodd}
\hat{\rm L}_M(\Pi_{+}^{\odd})
&=&\u(\vec q)i\G_5u(\vec q')
{\lL\lS\over\MBS\dM}
\left[ \gs \exp\left(-{\MS^2\over\MBS}\right) \right. \nonumber\\
&& \left. \quad -\gl \exp\left(-{(\ML+\m)^2\over\MBS}\right)\right]
+({\rm Cont.}),
\end{eqnarray}
for the odd part, where we define $\dM\equiv\ML+\m-\MS$, 
$\gs\equiv g(\MS, \,\vec p^2=0)$ and $\gl\equiv g(\ML+\m, \,\vec p^2=0)$.
$({\rm Cont.})$ denotes the continuum contribution coming from the last term of \Eq.{Ansatz},
in which we employ the QCD duality assumption and replace it by the corresponding OPE of the correlation 
function at $|q_0| >\sqrt{s_{th}}$.

In Eqs. (\ref{dI}) and (\ref{dIodd}), one sees that 
the poles bring the coupling constant at two different kinematical points,
$\gs\equiv g(\MS, 0)$ and 
$\gl\equiv g(\ML+\m = \MS + \dM, 0)$, while
the one defined in \Eq.{gpiLS} is at 
the kinematical point, $q^2=\MS^2$, $q^{'2}=\ML^2$, i.e.,
\begin{eqnarray}
g=g\left(q_0=\MS,{\vec p}^2=-\m^2+{(\MS^2+\m^2-\ML^2)^2\over4\MS^2}\right).
\end{eqnarray} 
Therefore in principle we need an interpolation.
However, we expect that the differences are small
because $\m$ and $\dM$ are both small compared to the baryon
masses.  Thus we will regard $\gs$ in the sum rule as the coupling constant.

The sum rules are obtained by equating these Borel-transformed correlation functions 
with the corresponding OPE terms.
The OPE's of $\Pi_{+}^{\even}$ and $\Pi_{+}^{\odd}$ are rather lengthy and therefore 
we give the explicit forms in Appendix B. 
In order to evaluate $\gs$, we operate
$$ \hat{\rm P}_{\Sigma} \equiv (\ML+\m)^2 - \MBS {\partial\over \partial \MBS} \MBS   $$
on the both sides of the sum rule and eliminate the $\gl$ terms.
Then we obtain
\begin{eqnarray}
\gs \,{\MS(\MS+\ML+\m) \over\MBS} \exp\left(-{\MS^2\over\MBS}\right)  
&=& \hat{\rm P}_{\Sigma} \left[{\bar\Pi_{+}^{\even}(M) \over \lL\lS} \right] ,
\label{eq:E_SR}\\
\gs \,{\MS+\ML+\m\over\MBS} \exp\left(-{\MS^2\over\MBS}\right)  
&=& \hat{\rm P}_{\Sigma} \left[{\bar\Pi_{+}^{\odd}(M) \over \lL\lS} \right]
\label{eq:O_SR}
\end{eqnarray}
with $\bar\Pi_{+}^{\even/\odd}(M) $ given in Appendix B.
The unknown constants, $\lL$ and $\lS$, are calculated from 
the vacuum-to-vacuum sum rule (mass sum rule) with the same interpolating fields and 
the Borel mass, whose explicit forms are also given in Appendix B.

In numerical analyses, the threshold parameters, 
$\sqrt{s_{th}}$ for $\Pi_+$, and 
$\sqrt{s_0}$ for the $\Lambda$ and $\Sigma$ mass sum rules, given in
Eqs.~(\ref{eq:Lam_SR}) and (\ref{eq:Sig_SR}), 
are chosen to be equal, 
because the continuum mainly comes from the $S=-1$ excited baryons 
and therefore is expected to be common to $\Pi_+$ and the mass sum rules.

Two sum rules, Eqs.~(\ref{eq:E_SR}) and (\ref{eq:O_SR}), should in principle give
the same result.
However, we have seen in sect.~\ref{sec:mbb-coupling} that the results depend
on the choice of the Dirac structure and that the tensor (T) sum rule 
is most reliable because it contains less contribution from the continuum and has
weaker dependence on the choice of the interpolating field.
In the present off-diagonal case, we employ the same criteria.
In fact, from the structure of the $\bar\Pi_{+}^{\even/\odd}$,
we find that $\Pi_{+}^{\odd}$
is superior to $\Pi_{+}^{\even}$ because $\Pi_{+}^{\odd}$ contains mainly the T structure
and therefore is less dependent on the continuum.
Actually, it is easy to check that $\Pi_{+}^{\odd}$ reduces to the OPE of the T structure
in the chiral limit.
In order to quantify this statement, we check how large the continuum
contribution is in each sum rule.
A numerical analysis tells us that (in the SU(3) limit) about 40\% of the continuum
contribution comes from the region above the threshold for the $\Pi_{+}^{\even}$
sum rule, Eq.~(\ref{eq:E_SR}), while it is less than 16\% for the $\Pi_{+}^{\odd}$
sum rule, Eq.~(\ref{eq:O_SR}).
Therefore we conclude that $\Pi_{+}^{\odd}$ sum rule is more reliable.

Another advantage of the $\Pi_{+}^{\odd}$ sum rule is that the main term in OPE
is proportional to the $\qq$ condensate, which is divided out by the main term of 
the baryon mass sum rule, This elimination of the $\qq$ condensate reduces 
ambiguity in the numerical results.
Thus we employ the $\Pi_{+}^{\odd}$ sum rule in the following analysis.

\subsection{Results}

We evaluate the $\pi\Lambda\Sigma$ coupling constant 
both in the SU(3) limit and in the realistic broken SU(3) case.
In getting the final values, we choose the parameter according to Table \ref{tab:table3},
except that in the SU(3) limit, we make $m_s=m_u \sim 0$ and $\braket{\bar s s} = \braket{\bar u u}$.
When we take the SU(3) limit, we use the nucleon mass for both the $\Lambda$ and $\Sigma$ masses,
while in the broken SU(3) calculation, the observed masses of the baryons are used.
To get the final results, we have to choose the threshold parameter and 
the Borel mass window.  In the SU(3) limit, the choice given in sect.~\ref{subsec:F/D} is taken, that is,
$2.07\leq s_{th} = s_0 \leq 2.57\ {\rm GeV}^2$ 
and 
$0.65 \leq M^2 \leq 1.24\ {\rm GeV^2}$.
In the broken SU(3) case, we set the threshold to above the first 
excited state of $\Sigma$ resonance, 
$2.76\leq s_{th} = s_0 \leq 3.27\ {\rm GeV}^2$ 
($\sqrt{s_{th}} = \sqrt{s_0} \sim 1.66-1.81$ GeV)
and consider the Borel window,
$1.00 \leq M^2 \leq 1.69\ {\rm GeV^2}$.

\begin{table}[htbp]
\caption{\label{tab:table3}
QCD parameters. We always assume that $\braket{\bar d d}= \braket{\bar u u}$, and 
employ the same $m_0^2 \equiv 
{\qGq \over\qq}$ for the $u$, $d$ and $s$ quarks.
The OPE terms which are proportional to $m_u$ or $m_d$ are neglected, that is 
equivalent to the choice $m_u=m_d=0$.}
\begin{ruledtabular}
\begin{tabular}{cccccc}
$\braket{\bar u u}$  & $\braket{\bar s s}/\braket{\bar u u}$  & $m_s$ & $m_0^2$
& $\delta^2$ & $\GG$\\
\hline
$ -(0.23~ {\rm GeV})^3$ & 0.8 & $0.12 {\rm GeV}$
& $0.8~{\rm GeV}^2$ & $0.2~{\rm GeV}^2$ & $(0.33~{\rm GeV})^4$  \\
\end{tabular}
\end{ruledtabular}
\end{table}

%%%%%%%%%  figure %%%%%%%%%%%%%%%
\begin{figure}[htbp]
\caption {$g_{\pi\Lambda\Sigma}$ from the $\Pi_{+}^{\odd}$ sum rule plotted against the 
squared Borel mass $\MBS$.
Thin solid (dashed) line denotes $g_{\pi\Lambda\Sigma}$ in the SU(3) limit
using the continuum threshold $s_{th} = s_0 = 2.07$ (2.57) ${\rm GeV}^2$.
Thick solid (dashed) line denotes $g_{\pi\Lambda\Sigma}$ 
including the SU(3) breaking effects 
using $s_{th} = s_0 = 2.76$ (3.26) ${\rm GeV}^2$.
}
\begin{center}
\includegraphics*[width=10cm]{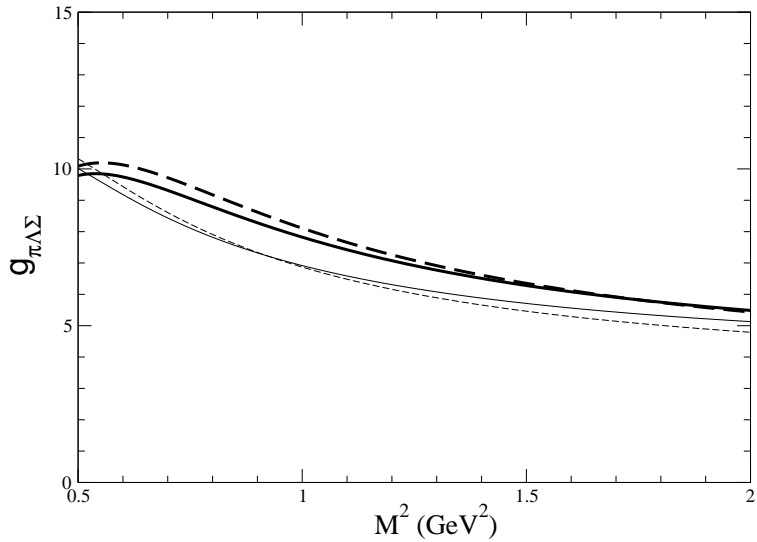}
\end{center}
\label{Borel_curve}
\end{figure}
%%%%%%%%%%%%%%%%%%%%%%%%%%%%%%%%

Fig.~\ref{Borel_curve} gives the resulting Borel curves for the SU(3) limit and
for the broken SU(3) case.
Comparing the solid lines with the corresponding dashed lines,
we confirm that the results are not sensitive to the choice 
of the continuum threshold.

The final results can be read off from the Borel curves by averaging over the values 
inside a Borel window.
One sees that the curves for the SU(3) limit are slightly lower than those in the broken SU(3) case.
We, however, have to shift the Borel window as given above so as to take account of
the larger $\Lambda$ and $\Sigma$ masses.  
We use the Borel window $0.65 \leq \MBS \leq 1.24 {\rm GeV}^2$
in the SU(3) limit and $1.00 \leq \MBS \leq 1.69 {\rm GeV}^2$ in the broken SU(3) case.
In the end, the obtained values of 
the coupling constants have little difference.
Then we obtain
\begin{eqnarray}
 &&g_{\pi\Lambda\Sigma}(\hbox{SU(3) limit}) =  7.5 \pm 1.3 ,
	 \nonumber\\
 &&g_{\pi\Lambda\Sigma}(\hbox{broken SU(3)}) =   6.9 \pm 1.0.
     \nonumber
\end{eqnarray}
As these numbers are consistent with each other, we conclude that the effect of 
the SU(3) breaking is weak in the $\pi\Lambda\Sigma$ coupling.

In the SU(3) limit the coupling constant, $g_{\pi\Lambda\Sigma}$, must be related 
to the $g_{\pi N}$ through the relation given in terms of the $F/D$ ratio, Eq.~(\ref{eq:su3-deSwart}),
$$  g_{\pi\Lambda\Sigma} = {2\over \sqrt{3}} (1-\alpha ) g_{\pi N}  .  $$
In order to see this relation, we apply the sum rule derived from the projected correlation function 
to $g_{\pi N}$.
Using the same parameters, the continuum threshold and the Borel mass window,
we obtain $g_{\pi N} \sim 9.6 \pm 1.6$, which corresponds to $F/D=0.47$ ($\alpha = 0.32$).
This value is smaller than $F/D = 0.65\pm 0.10$ determined from the T sum rule in sect.~\ref{subsec:F/D}.
This discrepancy is attributed mainly to the ambiguity brought by the mass sum rule, by which the
correlation function is divided in order to eliminate the factor $\LL$ or $\lambda_N^2$.
Generally the mass sum rule has moderate Borel mass dependence and therefore causes some
ambiguity.
Note that the $F/D$ ratio is obtained without using the mass sum rule in sect.~\ref{subsec:F/D}.
A further discrepancy may come from the difference in the method of extracting the coupling
constant.  Sum rules constructed by the derivative with respect to the Borel mass 
are in general less trustable than the global Borel mass fit.
Another source of discrepancy is the difference in the choice of the interpolating field.
As was pointed out in sect.~\ref{subsec:F/D}, the $F/D$ ratio from the interpolating field
\`a la Ioffe ($t=-1$) tends to be underestimated.  We indeed see in Fig.~\ref{fig:F/D-T} 
that the $F/D$ ratio at $t=-1$ is rather below the central value.
We also suspect that the $F/D$ ratio from the projected sum rule contains some spurious $t$-dependence 
because we do not take the chiral limit and then the OPE has contributions from the PS and PV structure.
Thus we conclude that the $F/D$ ratio obtained in the T sum rule is more reliable than that given here.

The value of $g_{\pi N}$ is consistent with the previous sum rule calculations,%
\footnote{The calculations done with the wrong sign of $\delta^2$ tend to give a larger value 
for $g_{\pi N}$ \cite{BK1,KD1,KM2}. See Appendix A for details.}
but it is smaller than the value determined from the strength of the one-pion exchange potential.
This discrepancy is still an open problem.

Earlier, an attempt was made to compute the $g_{\pi\Lambda\Sigma}$ coupling constant using the three-point
correlation function by Choe \cite{Choe}, giving $g_{\pi\Lambda\Sigma}\sim 10.79$, a larger 
number than our result.
It should, however, be noted that the $p^2\to 0$ limit is taken to evaluate the coupling constant, which, 
as we stressed in sect.~\ref{sec:formalism}, is not allowed in computing the OPE.

Recently, Lutz and Kolomeitsev \cite{Lutz} performed an extensive analysis of experimental data 
using relativistic chiral SU(3) formulation.
They have obtained the coupling constants from fitting to the scatterings data and obtain, among others,
$g_{\pi\Lambda\Sigma} \sim 10.4$, which is somewhat larger than our prediction.  
On the other hand, a recent coupled-channel analysis of the $\bar K N$ scattering has predicted 
much smaller value \cite{Keil}.
The obtained value, $g_{\pi\Lambda\Sigma}/g_{\pi N} \sim 0.4$, suggests a large SU(3) violation,
while our result is consistent with SU(3) symmetry with $\alpha \sim 0.4$.

\section {Conclusion}
\label{sec:conclusion}

As the fundamental theory of the strong interaction, QCD is to be applied directly to the
hadronic interactions.  We have presented a part of such attempts that the meson-baryon
coupling constants are calculated in the QCD sum rule approach.
It is shown that the coupling constants
are expressed in terms of the nonperturbative QCD parameters, such as the quark condensates, 
gluon condensates, as well as the pion (light-cone) wave functions.

We have addressed several technical issues in this report.
(1) It is advantageous to take the two-point correlation function of baryons to derive
sum rules for the coupling constants.  The correlation function
is evaluated for the initial meson
state and the final vacuum so that the coupling vertex appears in the middle.
(2)  We have considered the analytic structures of the correlation function and have
pointed out that the double pole term, one pole from the initial baryon and the other from the final
baryon, is carefully taken out to evaluate the coupling constant.
(3) In the case of the baryon-diagonal couplings, the double pole is easily extracted by the
Borel transform, and we have found that the most appropriate Dirac structure in the two
point function is the tensor type, proportional to $\gamma^5\sigma_{\mu\nu}q^{\mu}p^{\nu}$. 
Especially, this choice avoids spurious dependence on the choice of the interpolating
field operator for the octet baryons.
(4) In order to derive workable sum rule for the baryon-off-diagonal couplings, we
have found that the projected correlation function is most useful.
Among the two sum rules, even (E) and odd (O), we have found that the odd one gives 
the most reliable result.

We have emphasized that the SU(3) symmetry must be recovered if we turn off the
quark mass differences.  Then the coupling constants of the SU(3) octet mesons and
octet baryons can be expressed in terms of two constants, overall constant, represented
by $g_{\pi N}$ for instance, and the $F/D$ ratio.  Because the $F/D$ ratio is a completely
free parameter not determined by the SU(3) symmetry, it is important and interesting to 
determine this ratio directly from QCD.  
We have found that the sum rule in the SU(3) limit is fully consistent with SU(3) and 
gives 
$F/D = 0.65 \pm 0.10$, which almost coincides with 
the value derived from the quark model, or the SU(6) spin-flavor symmetry,
$F/D = {2\over 3}$, or $\alpha\equiv {F\over F+D}=0.4$.
This is also consistent with the $F/D$ ratio of the axialvector coupling constants
obtained from the beta decay rates of the hyperons.

We have calculated the $\pi\Lambda\Sigma$ coupling constant as an example of the
baryon-off-diagonal coupling.
We have found that the effect of the SU(3) breaking is small by comparing the
results in the SU(3) limit, $g_{\pi\Lambda\Sigma} = 7.5\pm 1.3$, and 
in the broken SU(3) case, $g_{\pi\Lambda\Sigma} = 6.9\pm 1.0$.
In fact, even at the level of the OPE, it can be seen that effects of the SU(3) breaking are weak 
as far as the pion-baryon couplings are concerned.  
The SU(3) breaking is taken into account in the QCD sum rule as
the quark mass term, $m_s \ne m_u$ and indirectly in terms of the
difference in the quark condensates, $\sbs \ne \uu$
and $\sGs\ne\uGu$.
However, both of these contributions do not appear in the leading terms of OPE
for the pion-baryon couplings.  Thus we see that not only the $\pi\Lambda\Sigma$
coupling, but also the other $\pi$-baryon couplings, like $\pi\Sigma\Sigma$, and
$\pi\Xi\Xi$, may not deviate much from the SU(3) values.

It is, however, not the case when we consider the couplings of the $\eta$, and maybe $K$, 
mesons.  The sum rule involves the meson mass, $m_{\eta,K}$ and the decay constants, $f_{\eta,K}$,
in the leading terms.  Therefore the SU(3) breaking of order a few tens of per cent can 
easily predicted in the QCD sum rule \cite{KD1}. 

In future analyses, it is desirable to calculate the $K B_1 B_2$ couplings,
such as $KN\Lambda$ and $KN\Sigma$,  which 
are phenomenologically very important, for instance, in the hypernuclear physics.%
\footnote{There have been several works on the QCD sum rules for the $K N\Lambda$ and $K N\Sigma$ couplings \cite{KNL_coupling}.
All but the first one, however, employ three-point correlation functions and therefore are not consistent with OPE.
Also their results do not agree with each other.}

\section*{Acknowledgment}
A part of this work is done in collaboration with Drs.~Osamu Morimatsu, Su Hong Lee and Hungchong Kim.
We would like to thank them for fruitful collaborations and stimulating discussions. 
This work is in part supported financially by the Grant for Scientific Research, No. 11640261 and 13011533,
from the Ministry of Education, Culture, Science and Technology, Japan.  T.D. acknowledges the support 
of the Japan Society for the Promotion of Science (JSPS).

\appendix
\def\px{p\cdot x}

\section{}

We calculate the Wilson coefficients of the short-distance expansion in two steps:
we perform the light-cone expansion of the correlation function first and the short-distance 
expansion of the light-cone operators second.
The reason for doing this is to use the parametrization of the vacuum-to-pion matrix elements 
of the light-cone operators given in Ref. \cite{Belyaev},
\begin{eqnarray}
\braket{0|\bar{u}i\gamma_5 u|\pi^0(p)} &=&
- \frac{\uu}{f_\pi} + i (p\cdot x) \frac{\uu}{2f_\pi} +
(p\cdot x)^2 \frac{\uu}{2f_\pi}  ( \frac{1}{3}+\frac{1}{30}B_2 ) \nonumber\\
&& + {\cal O}( (p\cdot x)^3 ) \\
\braket{0|\bar{u}\gamma_\mu \gamma_5 u|\pi^0(p)} 
&=&
if_\pi p_\mu + \frac{1}{2} f_\pi (p\cdot x)p_\mu \nn
&&
-\frac{1}{18} i f_\pi \delta^2 (p\cdot x) x_\mu 
+ \frac{5}{36} if_\pi \delta^2 x^2 p_\mu  
%- \frac{3}{20} if_\pi (1+\frac{2}{7}a_2) (p\cdot x)^2 p_\mu 
\nn
&& 
+ \frac{5}{72} f_\pi \delta^2  (p\cdot x) x^2 p_\mu
- \frac{1}{36} f_\pi \delta^2 (p\cdot x)^2 x_\mu 
\nn
&& 
+ {\cal O}(p^3 x^3) + {\cal O}((p)^0 x^4)
\end{eqnarray}
\begin{eqnarray}
\braket{0|\bar{u}\gamma_5 \sigma_{\mu\nu} u|\pi^0(p)} 
&=&
-i (p_\mu x_\nu - p_\nu x_\mu ) 
\frac{\uu }{6 f_\pi} \left( 1 - \frac{1}{2} i (p\cdot x) \right) + {\cal O}( (p\cdot x)^3)
\end{eqnarray}
%%%%%%%%%%%%%%%%%%%%%%%%%%%%%%%%%%%
%The contribution from pion 3-particle WF
\begin{eqnarray}
\lefteqn{
\braket{0|u^a (x) g_s G^A_{\mu\nu}(vx) \bar{u}^b(0)|\pi^0(p)}
} \nn
&=&
+ i\frac{f_{3\pi}}{16\sqrt{2}} t^A_{ab} 
\left[ \gamma_5 \sigma_{\lambda\mu} p^\lambda p_\nu 
- \gamma_5 \sigma_{\lambda\nu} p^\lambda p_\mu \right] \nn
&& 
+ \frac{if_\pi \delta^2}{252\cdot 16} t^A_{ab}
\Big[\ +(7+6\epsilon)\ (\gamma_5 \gamma_\mu p_\nu -  \gamma_5 \gamma_\nu p_\mu )\ (p\cdot x)  \nn
&& \qquad \qquad \qquad \quad
+(7+18\epsilon)\ \gamma_5 \fslash{p}\ ( p_\mu x_\nu -  p_\nu x_\mu )
\ \Big] \nn
&& - \frac{1}{16}\ t^A_{ab}\ \epsilon_{\mu\nu}^{\ \ \ \alpha\beta} \nn
&& \qquad \times
\Big[ +i f_\pi \delta^2\ \gamma_{\alpha} p_{\beta}\ 
\left\{ \frac{1}{3} - \frac{i}{84}(p\cdot x)(7+6\epsilon - 2v(-7+6\epsilon) )
\right\}  \nn
&& \qquad \qquad
+ f_\pi \delta^2\ \fslash{p}\ p_{\alpha} x_{\beta}\ 
\left\{ \frac{1}{252}(-7+6\epsilon) + \frac{1}{126} v (7-6\epsilon)
\right\}
\ \Big]   
\end{eqnarray}
Here $t^A\equiv {\lambda^A\over 2}$ is the color SU(3) generator, 
$g_s$ is the renormalized coupling constant of the QCD
and $G_{\rho\sigma}^A$ is the gluon field tensor.
We also define
$G_{\alpha\beta}\equiv \sum_A G_{\alpha\beta}^A t^A$
and $\tilde G_{\alpha\beta}\equiv{1\over2}\epsilon_{\alpha\beta\mu\nu}G^{\mu\nu}$
with $\epsilon_{0123}=-\epsilon^{0123}=1$.

Several parameters arise in the matrix elements:
$f_\pi$ ($=93$ MeV) is the pion decay constant, 
$f_{3\pi}$ ($=0.0035 ({\rm GeV})^2$) and $B_2 \equiv -{30\over\sqrt{2}}  {f_{3\pi}f_{\pi}\over\uu}$
($=0.57$) are the parameters coming from the twist-3 pion wave function,
and $\epsilon$ ( $= 0.5$) is from the twist-4 pion wave functions \cite{Belyaev}.
We choose 1 GeV for the renormalization scale.

The $\delta^2$ is another parameter from the twist-4 pion wave function 
defined, according to Novikov et al.~\cite{Novikov}, by 
\begin{equation}\label{eq:delta}
\langle0|\bar d g_s \tilde G_{\mu\nu}\gamma^\nu u|\pi^+(p)\rangle
=\sqrt{2}i f_\pi \delta^2 p_\mu \, .
\end{equation}
In previous calculations, however, there was some confusion on the sign of this constant.
Therefore, we here estimate $\delta^2$ in our notation according to Ref.~\cite{CZ}.

Let us consider the correlation function
\begin{eqnarray}
T_\mu=i\int dxe^{iqx}\<>{0|T[\d(x)\G_\mu\G_5g_s\sigma^{\alpha\beta}
G_{\alpha\beta} u(x)\u(0)\G_5d(0)]|0}.
\end{eqnarray}
Using the formula
$\<>{0|:q^a\bar q^bg_sG_{\alpha\beta}^h:|0}t^h_{ba}
=-{1\over48}\<>{0|\bar q g_s\sigma^{\mu\nu}G_{\mu\nu} q|0}\sigma_{\alpha\beta}$,
%$\sigma_{\alpha\beta}\sigma^{\alpha\beta}=12$, and $\sigma_{\alpha\beta}\G_\mu\sigma^{\alpha\beta}=0$,
we calculate the OPE of the correlation function, which gives to the lowest dimension 
\begin{eqnarray}\label{OPE}
T_\mu&=&\<>{\bar u g_s\sigma^{\alpha\beta}G_{\alpha\beta} u}\, {q_\mu\over q^2}.
%-{2\over3}\pi^2 \,\uu\,\<>{{\alpha_s\over\pi}G^2}\,{q_\mu\over q^4}.
\end{eqnarray}
The spectral function of the correlation function is given by
\begin{eqnarray}\label{Spectral}
{1\over\pi}{\rm Im}T_\mu
&=&(2\pi)^3\sum_n\left\{\delta^4(p-q)
\<>{0|\d\G_\mu\G_5{g_s\sigma_{\alpha\beta}G^{\alpha\beta}} u|n\rangle\langle n|\u\G_5d|0}\right.
\cr&&+\delta^4(p+q)
\left.\<>{0|\u\G_5d|n\rangle\langle n|\d\G_\mu\G_5
{g_s\sigma_{\alpha\beta}G^{\alpha\beta}} u|0}\right\}
\cr&=&(2\pi)^3\int{d^3p\over(2\pi)^32\omega_p}\left\{\delta^4(p-q)
\<>{0|\d\G_\mu\G_5
{g_s\sigma_{\alpha\beta}G^{\alpha\beta}} u|\pi^+(p)\rangle\langle\pi^+(p)|\u\G_5d|0}\right.
\cr&&+\left.\delta^4(p+q)
\<>{0|\u\G_5d|\pi^-(p)\rangle\langle\pi^-(p)|\d\G_\mu\G_5
{g_s\sigma_{\alpha\beta}G^{\alpha\beta}} u|0}\right\}
+ \hbox{(non-pion pole terms)}
%\cr&=&{1\over2\omega_p}[\delta(q_0-\omega_p)+\delta(q_0+\omega_p)]
%\left(i\sqrt2f_\pi A\right)
%\left(-\sqrt2if_\pi{m_\pi^2\over m_u+m_d}\right)
\cr&=&\delta(q^2-m_\pi^2)q_\mu{2f_\pi^2m_\pi^2\over m_u+m_d}A
+ \hbox{(non-pion pole terms)} .
\end{eqnarray}
The constant $A$ is defined by
\begin{eqnarray}\label{Matele}
\<>{0|\d\G_\mu\G_5g_s\sigma^{\alpha\beta}G_{\alpha\beta}u|\pi^+(p)}
&\equiv& i\sqrt2f_\pi p_\mu A,\\
\<>{\pi^-(p)|\d\G_\mu\G_5
{g_s\sigma_{\alpha\beta}G^{\alpha\beta}} u|0} &\simeq& -i\sqrt2f_\pi p_\mu A,
\end{eqnarray}
where we use the fact that 
$\<>{\pi^-(p)|\d ig_s{G_\mu}^\beta\G_\beta\G_5 u|0}$ is of higher order in
the chiral expansion \cite{Novikov}.
The phases of the pion states are taken as
\begin{eqnarray}
\<>{0|\d \G_\mu \G_5 u|\pi^+(p)}&=& i\sqrt2 f_\pi p_\mu \\
\<>{0|\u \G_\mu \G_5 d|\pi^-(p)}&=& i\sqrt2 f_\pi p_\mu .
\end{eqnarray}

Substituting \Eq.{OPE} and \Eq.{Spectral} for the left-hand and 
the right-hand sides of the dispersion relation as follows
\begin{eqnarray}
T(q^2)=-{1\over\pi}\int_0^\infty ds{{\rm Im}T(s)\over q^2-s},
\end{eqnarray}
respectively, at $Q^2=-q^2\rightarrow\infty$ we obtain 
\begin{eqnarray}
\<>{\bar u g_s\sigma^{\alpha\beta}G_{\alpha\beta}  u}\, {q_\mu\over Q^2}
=-{2f_\pi^2m_\pi^2\over m_u+m_d}\, {q_\mu A\over Q^2+m_\pi^2} .
\end{eqnarray}
Taking the chiral limit and using the relation 
$f_\pi^2m_\pi^2=-(m_u+m_d)\uu$, we find
\begin{eqnarray}
A={\<>{\bar u g_s\sigma^{\alpha\beta}G_{\alpha\beta}  u}
\over2\uu}={m_0^2\over2} .
\end{eqnarray}
Using the identity, 
$\G_\mu\sigma_{\alpha\beta}=i(g_{\mu\alpha}\G_\beta-g_{\mu\beta}\G_\alpha)
+\epsilon_{\mu\alpha\beta\nu}\G^{\nu}\G_5$,
one can rewrite the matrix element (\ref{Matele}) as
\begin{eqnarray}
\<>{0|\d g_s\G_\mu\G_5\sigma_{\alpha\beta}G^{\alpha\beta}u|\pi^+(p)}
%&=&\<>{0|\d g_s(i(g_{\mu\alpha}\G_\beta-g_{\mu\beta}\G_\alpha)
%+\epsilon_{\mu\alpha\beta\lambda}\G^\lambda\G_5)G^{\alpha\beta}\G_5u|\pi(p)}\cr
&=&2\<>{0|\d ig_s{G_\mu}^\beta\G_\beta\G_5u|\pi^+(p)}
+2\<>{0|\d g_s\tilde G_{\mu\nu}\G^{\nu} u|\pi^+(p)}.
\end{eqnarray}
The first term in this expression is of higher order in the chiral expansion \cite{Novikov}.
>From Eqs.~(\ref{eq:delta}) and (\ref{Matele}), we obtain
\begin{eqnarray}
\delta^2={m_0^2\over4}\sim 0.2\;{\rm GeV}^2.
\end{eqnarray}

The sign of $\delta^2$ is therefore determined by that of $m_0^2$, which has been 
relatively well studied as a (higher dimensional) chiral order parameter.
%Recent quenched lattice QCD calculation gives $m_0^2 \simgeq 1 ({\rm GeV})^2$ \cite{Doi_LQCD}.
In the analysis in ref.~\cite{BK1}, the $\delta^2$ with the opposite sign was used, which was
transferred to some of the subsequent studies \cite{KD1,DK1,KM2,DK2}.  
The wrong sign of $\delta^2$ happens to give
a larger value of the $\pi NN$ coupling constant, which tends to agree with experimental data.
In fact, if we use the correct $\delta^2$, then the coupling constants are reduced by about
30 \% or so and the agreement with data is somewhat spoiled.

\section{}

The operator product expansion (OPE) sides of the 
$\Pi_{+}^{\even}$ and $\Pi_{+}^{\odd}$ defined in Eq.~(\ref{eq:Pievenodd}) 
for the $\pi\Lambda\Sigma$ coupling constant are explicitly given in this Appendix. 

We define the correlation functions by
\begin{eqnarray}
\Pi (q,p ) &\equiv&
 i \int d^4x e^{iqx} 
\braket{0|{\rm T}[J_{\Sigma^0}(x)\ \bar{J}_{\Lambda}(0)]|\pi^0(p)} ,\\
\Pi_+(q,p) &\equiv&
\bar{u}_\Sigma (\bec{q})\gamma_0 \Pi \gamma_0 u_\Lambda(\bec{q}-\bec{p}) ,\\
\Pi_+^{\even}(q_0) &\equiv& \frac{1}{2} \left[ \Pi_+(q_0) + \Pi_+(-q_0) \right] ,\\
\Pi_+^{\odd} (q_0) &\equiv& \frac{1}{2q_0} \left[ \Pi_+(q_0) - \Pi_+(-q_0) \right]  ,
\end{eqnarray}
where the baryon interpolating fields are chosen according to Ioffe \cite{Ioffe} 
(that is, $t=-1$) as
\begin{eqnarray}
J_{\Lambda}(x) &=& \sqrt{\frac{2}{3}}  \epsilon_{abc} 
\left( 
 [u_a^T C\gamma_\mu s_b ] \gamma_5 \gamma^\mu d_c
-  [d_a^T C\gamma_\mu s_b ] \gamma_5 \gamma^\mu u_c 
\right) , \\
J_{\Sigma^0}(x) &=& \sqrt{2} \epsilon_{abc} 
\left( 
  [u_a^T C\gamma_\mu s_b ] \gamma_5 \gamma^\mu d_c
+ [d_a^T C\gamma_\mu s_b ] \gamma_5 \gamma^\mu u_c 
\right) .
\end{eqnarray}

We subtract the continuum contribution from the $\Pi_+^{\even/\odd}$, 
assuming the threshold parameter, $\sth$,
and then apply the Borel transform
\begin{eqnarray}
       \hat{\rm L}_M\equiv \lim_{{n\rightarrow\infty \atop -q_0^2\rightarrow\infty}
       \atop -q_0^2/n = \MBS}
       {(q_0^2)^n\over(n-1)!}\left(-{d\over dq_0^2}\right)^n ,
\end{eqnarray}
where $\MB$ is the Borel mass.

Finally, we obtain 
{\footnotesize
\begin{eqnarray}
\lefteqn{\bar{\Pi}^{\rm E}_+ (M^2) \equiv} \nonumber\\
&& 
\frac{4}{\sqrt{3}}\Biggl\{
M^2 C_2(\sqrt{s_{th}})  %%%%%%%%%%%%%%%
\biggl[
 \frac{1}{8\pi^2} \coeffa % PS
+ \frac{7}{48\pi^2} f_\pi ( E_\Lambda + m_\Lambda - \omega_p ) % PV
- \frac{1}{24\pi^2} f_\pi \omega_p  % PV2
\biggr] \nn
&& \quad
+ C_1(\sqrt{s_{th}})  %%%%%%%%%%%%%%%
\biggl[\ 
\left(
- \frac{1}{12\pi^2} f_\pi \delta^2          % PV
+ \frac{1}{16\pi^2} \coeffa (m_q - 2 m_s)   % PV
\right) ( E_\Lambda + m_\Lambda - \omega_p ) 
\biggr] \nn
&& \quad 
+ \frac{1}{M^2}    %%%%%%%%%%%%%%%%%
\biggl[\ 
-\frac{1}{48}\GG \coeffa % PS
+ \frac{1}{6} \coeffa (m_q \qq + m_s \sbs ) % PS
- \frac{2}{3} \coeffa (m_q \sbs + m_s \qq ) % PS
\nn
&& \qquad \qquad
+ \left(
- \frac{1}{6}\coeffa \{\qq - \frac{4}{3} \sbs \} % PV
+ \frac{5}{288}\GG f_\pi % PV
\right.
\nn
&& \quad \quad \quad \quad \quad
\left.
+ \frac{1}{4}f_\pi (m_q \qq + m_s \sbs ) % PV
- \frac{1}{3}f_\pi (m_q \sbs + m_s \qq ) % PV
\right) ( E_\Lambda + m_\Lambda - \omega_p ) 
\nn
&& \qquad \qquad
+ \left(
- \frac{1}{3} \coeffa \{ \qq - \frac{2}{3} \sbs \}  % PV2
+ \frac{1}{144} \GG f_\pi                          % PV2
\right) \omega_p 
- \frac{1}{3} f_\pi \qq \omega_p ( E_\Lambda + m_\Lambda ) % T
\biggr] \nn
&& \quad 
+ \frac{1}{M^4}    %%%%%%%%%%%%%%%%%
\biggl[\ 
- \frac{1}{12} \coeffa (m_q \sGs + m_s \qGq ) % PS
\nn
&& \qquad \qquad
+ \left(
 \frac{1}{48}\coeffa \{\qGq -\frac{2}{3}\sGs \} % PV
+ \frac{1}{48\cdot 54}\GG f_\pi \delta^2 % PV
- \frac{1}{108}f_\pi (m_q \qq + m_s \sbs ) \delta^2 % PV
\right.
\nn 
&& \quad \quad \quad \quad
\left.
- \frac{1}{27}f_\pi (m_q \sbs + m_s \qq ) \delta^2  % PV
- \frac{1}{144} f_\pi (m_q \qGq + m_s \sGs )       % PV
\right) ( E_\Lambda + m_\Lambda - \omega_p ) 
\nn
&& \qquad \qquad
+ \left(
 \frac{1}{12} \coeffa \{\qGq - \frac{2}{3}\sGs \} % PV2
+ \frac{1}{648} \GG f_\pi \delta^2                 % PV2
\right) \omega_p 
\nn
&& \qquad \qquad
+ \left(
 \frac{1}{27} f_\pi \delta^2 ( 15 \qq - 2 \sbs ) % T
+ \frac{1}{12}f_\pi \qGq                         % T
- \frac{1}{216}\GG \coeffa                       % T
\right) \omega_p ( E_\Lambda + m_\Lambda ) 
\biggr]
\ \ \Biggr\}
\label{SReven}
\end{eqnarray}%
}
and 
{\footnotesize
\begin{eqnarray}
\lefteqn{
\bar{\Pi}^{\rm O}_+ (M^2) \equiv
}
\nonumber\\
&& 
\frac{4}{\sqrt{3}}\Biggl\{
C_1(\sqrt{s_{th}})  %%%%%%%%%%%%%%%
\biggl[
- \frac{1}{8\pi^2} \coeffa \omega_p   % PS
- \frac{1}{6\pi^2} f_\pi \omega_p (E_\Lambda + m_\Lambda ) % PV
+ \frac{1}{8\pi^2} \coeffa (m_q - m_s ) % PV
\nn
&& \qquad \qquad \qquad
+ \left(
- \frac{1}{24\pi^2} \coeffa % T
+ \frac{1}{8\pi^2} f_\pi m_q % T
\right)  (E_\Lambda + m_\Lambda ) 
\biggr] \nn
&& \quad
+ \frac{1}{M^2}    %%%%%%%%%%%%%%%%
\biggl[
- \frac{1}{3} f_\pi \{\qq - \sbs \} \omega_p % PS
+ \frac{5}{72\pi^2} f_\pi \delta^2 \omega_p (E_\Lambda + m_\Lambda)  % PV
- \frac{1}{3} \coeffa \{ \qq - \sbs \} % PV2
\nn
&& \qquad \qquad
+ \left(
- \frac{1}{3} f_\pi \qq % T
- \frac{5}{36\pi^2} f_\pi \delta^2 m_q % T
\right) (E_\Lambda + m_\Lambda ) 
\biggr] \nn
&& \quad 
+ \frac{1}{M^4}    %%%%%%%%%%%%%%%%%
\biggl[\ 
\left(
 \frac{5}{18} f_\pi \{\qq - \sbs \}  \delta^2 % PS
+ \frac{1}{24} f_\pi \{\qGq - \sGs \}  % PS
+ \frac{1}{48}\GG \coeffa  % PS
\right) \omega_p
\nn
&& \qquad \qquad
+ \left(
 \frac{2}{3}\coeffa  (\frac{1}{3}+\frac{1}{30}B_2)
\{\qq - \sbs \}            % PV
- \frac{f_{3\pi}}{3\sqrt{2}} \{ \qq - \sbs \}   % PV
- \frac{1}{48}\GG f_\pi   % PV
\right)  \omega_p (E_\Lambda + m_\Lambda ) 
\nn
&& \qquad \qquad
+ \frac{1}{24} \coeffa \{\qGq - \sGs \} % PV2
\nn
&& \qquad \qquad
+ \left(
 \frac{13}{54}f_\pi \delta^2 \qq % T
+ \frac{1}{24}f_\pi \qGq  % T
- \frac{1}{432}\GG \coeffa  % T
\right. 
\nn
&& \qquad \qquad \qquad
\left.
- \frac{1}{18} \coeffa (m_q\qq + m_s\sbs ) % T
+ \frac{1}{9} \coeffa (m_q\sbs + m_s\qq )  % T
\right) (E_\Lambda + m_\Lambda ) 
\biggr]
\ \ \Biggr\} .
\label{SRodd}
\end{eqnarray}%
}
Here we use the label $q$ for the $u$ and $d$ quarks and define a function
\begin{eqnarray}
C_n(\omega)&=&1-\left[\sum_{k=1}^n{1\over(k-1)!}\left({\omega^2\over\MBS}\right)^{k-1}\right]
\exp\left(-{\omega^2\over\MBS}\right).
\end{eqnarray}

\medskip

The sum rules for the baryon masses
are given in terms of the two-point correlation function of the baryon interpolating field
operator.  We use the sum rules for the $\Lambda$ and $\Sigma$ masses
to eliminate the couplings of the interpolating fields, $\lL$ and $\lS$.
%The correlation function contains two terms, the even and odd terms.%
%\footnote{Note that the labels, ``even'' (\even), and ``odd'' (\odd), are reversed from the commonly used
%``chiral'' symmetry classification.  The present ``even'' sum rule are chiral odd one,
%which is regarded to give a better sum rule for the nucleon mass than the other sum rule. }
For the $\Lambda$ baryon, the sum rule calculated up to the dimension seven 
operators \cite{RRY,massSR} is given by
\begin{eqnarray}\label{SRLeven}
&&\ML{\lL^2\over \MBS}\exp\left(-{\ML^2\over\MBS}\right) \cr
&=&{1\over4\pi^4}\Bigg[\MBQ C_3(\sqrt{s_0})\Big\{{1\over12}(4m_q-m_s)\Big\}
       +\MBS C_2(\sqrt{s_0})\pi^2\Big\{-{4\over3}{\qq}+{1\over3}{\sbs} \Big\} \cr
&&\qquad +{1\over\MBS}\pi^4\Big\{{8\over9}m_q(-2{\qq}^2+13{\qq}{\sbs})
             +{8\over9}m_s(6{\qq}^2-2{\qq}{\sbs}) \cr
&&\qquad  +{1\over18}(4{\qq}-{\sbs})
    \<>{{\alpha_s\over\pi}G^{\alpha\beta}G_{\alpha\beta}}
                        \Big\}\Bigg] ,
\label{eq:Lam_SR}
%                                                \cr
%&\equiv&\bar\Pi_{\Lambda}^{\even}(M)
\end{eqnarray}
%and
%\begin{eqnarray}\label{SRLodd}
%&&{\lL^2\over \MBS}\exp\left(-{\ML^2\over\MBS}\right)\cr
%&=&{1\over4\pi^4}
%     \Bigg[M^4 C_3(\sqrt{s_0}){1\over8}
%   +C_1(\sqrt{s_0})\pi^2\Big\{
%    {1\over8}\<>{{\alpha_s\over\pi}G^{\alpha\beta}G_{\alpha\beta}}
%   +{4\over3}(2m_q-m_s){\qq} \cr
%& &\qquad +{1\over3}(3m_s-4m_q){\sbs}\Big\}
%   +{1\over\MBS}\pi^4\Big\{ {32\over9}{\qq}{\sbs}-{8\over9}{\qq}^2 \Big\} \Bigg]
%   \cr 
%&\equiv&\bar\Pi_{\Lambda}^{\odd}(M)
%\end{eqnarray}
where the continuum contribution is subtracted 
with $\sqrt{s_0}$ being the effective continuum threshold.

Similarly, for $\Sigma$, we obtain
\begin{eqnarray}\label{SRSeven}
&&\MS{\lS^2\over \MBS}\exp\left(-{\MS^2\over\MBS}\right) \cr
&=&{1\over4\pi^4}\Bigg[ 2\MBQ C_3(\sqrt{s_0})\Big\{{1\over8}m_s\Big\}
       +\MBS C_2(\sqrt{s_0})\pi^2\{-{\sbs}\}\cr
& &\qquad
   +{1\over\MBS}\pi^4\Big\{8m_q{\qq}{\sbs}+{16\over3}m_s{\qq}^2
    +{1\over6}{\sbs}\<>{{\alpha_s\over\pi}G^{\alpha\beta}G_{\alpha\beta}}
               \Big\}\Bigg] .
\label{eq:Sig_SR}
%                           \cr
%&\equiv&\bar\Pi_{\Sigma}^{\even}(M) 
\end{eqnarray}
%and
%\begin{eqnarray}\label{SRSodd}
%&&{\lS^2\over \MBS}\exp\left(-{\MS^2\over\MBS}\right)\cr
%&=&{1\over4\pi^4}\Bigg[\MBQ C_3(\sqrt{s_0}){1\over8}
%   +C_1(\sqrt{s_0})\pi^2\Big\{ m_s{\sbs}
%    +{1\over8}\<>{{\alpha_s\over\pi}G^{\alpha\beta}G_{\alpha\beta}}\Big\}
%   +{1\over M^2}\pi^4\Big\{{8\over3}{\qq}^2 \Big\}
%     \Bigg] 
%         \cr
%&\equiv&\bar\Pi_{\Sigma}^{\odd}(M) 
%.
%\end{eqnarray}
These are the chiral-odd sum rules, which are commonly acknowledged as to give a reliable sum
rule for the baryon mass.

\def \pnum(#1,#2,#3){{\bf{#1}}~%
\ifnum#2>1000 (#2) \else \ifnum#2>53 (19#2) \else (20#2) \fi\fi#3}

\def \NP(#1,#2,#3){{Nucl. Phys.} \pnum(#1,#2,#3)}
\def \PL(#1,#2,#3){{Phys.\ Lett.} \pnum(#1,#2,#3)}
\def \PRL(#1,#2,#3){{Phys.\ Rev.\ Lett.} \pnum(#1,#2,#3)}
\def \PRp(#1,#2,#3){{Phys.\ Rep.} \pnum(#1,#2,#3)}
\def \PR(#1,#2,#3){{Phys.\ Rev.}  \pnum(#1,#2,#3)}
\def \PTP(#1,#2,#3){{Prog.\ Theor.\ Phys.} \pnum(#1,#2,#3)}
\def \ibid(#1,#2,#3){{\it ibid.}\ \pnum(#1,#2,#3)}

\def\etal{{\it et al.}}

\end{document}